\documentclass[twocolumn, superscriptaddress, nofootinbib, amsmath,amssymb, aps, prd, floats, floatfix]{revtex4-2}

\usepackage{graphicx}
\usepackage{dcolumn}
\usepackage{bm}
\usepackage{multirow}
\usepackage{xcolor}
\usepackage{hyperref}
\usepackage{orcidlink}

\hypersetup{
    colorlinks=true,
    linkcolor=blue,
    filecolor=magenta,      
    citecolor=blue,
    urlcolor=blue,
    }

\def\apj{ApJ}
\def\apjl{ApJ}

\newcommand{\msun}{M$_{\odot}$}

\begin{document}

\title{A binary neutron star merger search pipeline powered by deep learning}

\author{Alistair McLeod\orcidlink{0000-0001-5424-8368}} 
\email{alistair.mcleod@research.uwa.edu.au}
\author{Damon Beveridge\orcidlink{0000-0002-1481-1993}}
\email{damon.beveridge@research.uwa.edu.au}
\author{Linqing Wen\orcidlink{0000-0001-7987-295X}}
\email{linqing.wen@uwa.edu.au}
\affiliation{Department of Physics, The University of Western Australia, 35 Stirling Hwy, Crawley, Western Australia 6009, Australia}

\author{Andreas Wicenec\orcidlink{0000-0002-1774-5653}}
\affiliation{International Centre for Radio Astronomy Research,
The University of Western Australia, 35 Stirling Hwy, Crawley, Western Australia 6009, Australia}

\date{\today}

\begin{abstract}

Gravitational waves are now routinely detected from compact binary mergers, with binary neutron star mergers being of note for multimessenger astronomy as they have been observed to produce electromagnetic counterparts. Novel search pipelines for these mergers could increase the combined search sensitivity, and could improve the ability to detect real gravitational wave signals in the presence of glitches and nonstationary detector noise. Deep learning has found success in other areas of gravitational wave data analysis, but a sensitive deep learning-based search for binary neutron star mergers has proven elusive due to their long signal length. In this work, we present a deep learning pipeline for detecting binary neutron star mergers. By training a convolutional neural network to detect binary neutron star mergers in the signal-to-noise ratio time series, we concentrate signal power into a shorter and more consistent timescale than strain-based methods, while also being able to train our network to be robust against glitches. We compare our pipeline's sensitivity to the three offline detection pipelines using injections in real gravitational wave data, and find that our pipeline has a comparable sensitivity to the current pipelines below the 1 per 2 months detection threshold. Furthermore, we find that our pipeline can increase the total number of binary neutron star detections by 12\% at a false alarm rate of 1 per 2 months. The pipeline is also able to successfully detect the two binary neutron star mergers detected so far by the LIGO-Virgo-KAGRA collaboration, GW170817 and GW190425, despite the loud glitch present in GW170817.

\end{abstract}

\maketitle

\section{Introduction}
\label{sec:bns_intro}

Gravitational waves (GW) from compact binary coalescences (CBCs) are now regularly detected by ground-based laser interferometers, with the LIGO \citep{ALIGO} and Virgo \citep{AVIRGO} interferometers detecting over 90 CBCs in the first three observing runs \citep{gwtc1, gwtc2, gwtc3}. Most of these CBCs are binary black hole (BBH) mergers, with only two confirmed binary neutron star (BNS) mergers \citep{170817,190425}, and two confirmed neutron star - black hole (NSBH) mergers by the end of the third observing run (O3) of the LIGO-Virgo-KAGRA collaboration. The first detected BNS merger, GW170817, is notable for its heralding of a new age of multimessenger astronomy with gravitational waves as a messenger \citep{170817, 170817_multimessenger}. A gamma-ray burst was serendipitously detected from the merger \citep{Goldstein2017}, as well as a kilonova and x-ray counterpart from follow-up observations \citep{170817_multimessenger, Haggard2017}. These observations provided a unique measurement of the Hubble constant \cite{GW_hubble}, and constraints on the neutron star equation of state \citep{Radice2018, Baiotti2019}. Further observations of BNS mergers could further constrain the Hubble constant and resolve the Hubble tension, and potentially reveal a link between BNS mergers and other transient signals, such as fast radio bursts. As the interferometers improve in sensitivity, and new interferometers such as KAGRA \citep{KAGRA} come online, the possibility of making more multimessenger detections warrants the development of new CBC pipelines to search for them.

CBCs are primarily detected by five search pipelines \citep{gstlal, pycbclive, mbta, Chu2022, cwb}, with four of the five pipelines using matched filtering to identify signals. The matched filtering pipelines use a bank of signal templates with unique intrinsic parameters to cover the mass-spin parameter space. These templates are cross-correlated with the incoming GW detector data to produce signal to noise ratio (SNR) time series. In the absence of noise, the highest SNR for an incoming GW signal is produced by the template with parameters that most closely match the true signal parameters. Triggers are produced when an SNR threshold is satisfied (for example an SNR $>$ 4 in one detector). These triggers are then clustered and assigned a significance using a ranking statistic. Ranking statistics typically take into account the peak SNR of the trigger, whether there are coincident triggers between the observing interferometers, and tests for signal consistency \citep{Allen2005}. Triggers are assigned a false alarm rate (FAR) based on a collected background of triggers, and triggers with a sufficiently low FAR are considered GW candidates. 

Despite the success of the current pipelines at detecting CBCs, it is worthwhile investigating new detection methods for several reasons. First, the overall search for CBCs benefits from multiple search pipelines using unique search methods \citep{DalCanton2021}. Unique search methods provide the possibility for the detection of events that would have been missed by other search methods, and joint detections with other searches provide supporting evidence that an event is a genuine CBC. Second, an area of ongoing research is the mitigation of non-Gaussian transient noise artefacts (glitches), which can produce high-SNR triggers that pipelines should avoid producing alerts on. A key challenge is the exclusion of glitches from analysis, without also excluding actual CBC signals from producing alerts \citep{170817}. As the detection rate of CBCs and the rate of instrumental glitches have both increased over time \citep{gwtc2, gwtc3}, CBC events contaminated by glitches will likely become more frequent in the future. A detection method that can successfully identify signals while minimizing the effect of glitches, as well as correctly identifying signals contaminated with glitches would be ideal. With these stipulations, a deep learning-based detection method is a logical choice for investigation \citep{Beveridge2024}.

Deep learning has found success as a useful tool for improving the accuracy or latency of several areas of gravitational wave data analysis (see e.g.~\citep{Cuoco2021, Dax2021, Gabbard2022, Chatterjee2023, Bahaadini2018, Essick2020, Skliris2022, Schafer2023, Nousi2023, Marx2024, Beveridge2024}). Gravitational wave strain-based BBH detection with deep learning has been shown to be effective, and potentially able to reach the sensitivity of the matched filtering detection pipelines in real detector noise \citep{Schafer2023,Nousi2023,Marx2024}. However, compared to BBH detection, there are additional challenges introduced when applying deep learning to lower mass signals like BNS mergers. Strain-based BNS detection methods face the issue that at current sensitivity, BNS mergers are present for $\mathcal{O}(100~\text{s})$ in detector data, meaning the signal power that accumulates at a detector is significantly more spread out compared to a BBH merger with equivalent SNR. A strain-based BNS detection method \citep{Krastev2020, Krastev2021, Schafer2020, Baltus2021, Qiu2023} has to either lose signal power by truncating the input window, or through other approximations made in preprocessing, which limits the achievable sensitivity. Spectrogram-based detection methods \citep{Wei2021, Aveiro2022} face the same issue. Consequently, a deep learning approach for BNS detection that can match the sensitivity of the matched filtering detection pipelines has yet to be demonstrated.

In this work, we investigate the use of a neural network (NN) based search pipeline for the detection of BNS mergers in the SNR time series produced by matched filtering. The advantage of detecting in the SNR time series is that CBC signal power is condensed relative to in the strain, which is especially beneficial for the longer-duration BNS mergers. SNR time series are also a readily available data product from the matched filtering pipelines, making online implementation relatively straightforward. This work is further motivated by \citep{Beveridge2024}, where we found that BBH detection with SNR time series produced promising sensitivity results, especially toward lower BBH masses. We train our NN on LIGO Hanford (H1) and LIGO Livingston (L1) real detector noise from the third observing run to ensure it is robust against real glitches. By characterizing our search pipeline's ranking statistic on past data and performing searches on the O3 injection set from the third gravitational wave transient catalog (GWTC-3) \citep{gwtc3}, we show that our NN can match the sensitivity of the current detection pipelines and can detect the two real BNS events, GW170817 and GW190425.

The structure of the remainder of this work is as follows. In Sec.~\ref{sec:dataset} we cover how we implement matched filtering, select our detector noise, and how we generate our template bank and training datasets. In Sec.~\ref{sec:NN} we cover the high-level architecture of our neural network and its training and validation. The method we use to run our search pipeline and assign false alarm rates is presented in Sec.~\ref{sec:bg}. In Sec.~\ref{sec:inj} we present the performance of our search pipeline using an injection set from the GWTC-3 offline analyses, as well as the pipeline's detection of the two real BNS events. In Sec.~\ref{sec:discussion} we summarize the findings of this work, discuss their implications, and discuss potential future improvements.

\section{Dataset Generation}
\label{sec:dataset}

In this section, we introduce the concept of matched filtering and discuss how we use matched filtering to generate our training and validation datasets. In Sec.~\ref{sec:MF} we define our implementation of matched filtering. Sec.~\ref{sec:template_bank} describes how we generate the BNS template bank used in the rest of this work. Sec.~\ref{sec:noise} covers how we acquire real noise for the training and validation datasets, as well as for our sensitivity tests. In Sec.~\ref{sec:dataset_construction} we create our training and validation datasets for our neural network. 

\subsection{Matched filtering}
\label{sec:MF}

Matched filtering is a signal processing technique commonly employed in gravitational wave research, as it is the optimal detection method for modeled signals in stationary Gaussian noise \citep{Allen2012}. It is the process of cross-correlating a signal template $s$ with incoming detector data $h$, and produces a signal-to-noise ratio (SNR) time series $\rho(t)$ \citep{Allen2012, Nitz2018}:

\begin{equation}
    \label{eqn:mf_SNR}
    \rho(t) = \frac{|z(t)|}{\sqrt{\langle s | s \rangle }}\,,
\end{equation}
\noindent
where $\langle s | s \rangle$ is the noise-weighted inner product of the template and $z(t)$ is the complex matched filter

\begin{equation}
    \label{eqn:SNR}
    z(t) = 4 \int^{f_{\text{high}}}_{f_{\text{low}}} \frac{\Tilde{s}(f) \Tilde{h}^*(f)}{S_n(f)} e^{2 \pi i f t} df\,,
\end{equation}

\noindent
where $S_n(f)$ is the estimated one-sided power spectral density (PSD) of the detector noise, $f_{low}$ and $f_{high}$ are the low and high frequency cutoffs used, and a tilde represents the Fourier transform of the template or data.

As a proposed input to a deep learning model, the SNR time series has a key benefit over detector strain data. In gravitational wave strain data, a CBC's signal power can be present for potentially hundreds of seconds, depending on the progenitor masses of the CBC and the low-frequency sensitivity of the interferometer. After matched filtering, however, the signal power is condensed into an SNR peak that is tens of milliseconds wide.

We implement matched filtering using Python's \textsc{NumPy} module \citep{Harris2020}, which allows us to do array-wise matched filtering to efficiently compute batches of SNR time series. We adapted our implementation of matched filtering from the \textsc{PyCBC} library \citep{pycbc_software}.

\subsection{Template bank generation}
\label{sec:template_bank}

We used \textsc{PyCBC}'s \texttt{pycbc\_geom\_aligned\_bank} method \citep{Brown2012,pycbc_software} to generate our template bank. This method uses the TaylorF2 metric \citep{Buonanno2009} to create a geometrical lattice of points which are then mapped to mass-spin values. These mass-spin pairs are the template parameters, which fill the parameter space to the desired coverage. This bank generation method is suitable as BNS signals are well-described by the inspiral-only TaylorF2 metric. Additionally, geometrical bank generation methods produce smaller template banks than equivalent stochastic methods \citep{Harry2009}, reducing our computational requirements. Another benefit of using this template bank generation method is that we can use the generated coordinate transformation matrices to approximate the overlap between templates and training waveforms, which speeds up training dataset construction. 
 
The input parameters for generating the template bank are shown in Table~\ref{tab:template_bank}. We set the maximum z-aligned spin magnitude S$_z$ to 0.05, as it is unlikely neutron stars would merge with a larger spin \citep{Brown2012, Zhu2018}. We generate templates with a maximum component mass of 3 \msun\, to ensure full coverage of BNS signals, even though we do not generate any training set signals with component masses above 2.6 \msun. This is because in training we generate signals in the source frame rather than in the detector frame, so the 3 \msun\,upper limit is necessary to compensate for high mass BNS systems being redshifted. The bank was generated with a minimum overlap of 0.98, meaning the maximum SNR loss due to template mismatch is 2\%. This set of parameters yields a bank of 30,858 templates. When we generate the templates for matched filtering, we generate them in frequency space using the TaylorF2 approximant.

\begin{table}[h]
    \centering
    \begin{tabular}{l @{\hskip 0.5cm} l}
         \hline
         \hline
         Parameter& Value  \\
         \hline

         Minimum component mass & 1 M$_{\odot}$\\
         Maximum component mass & 3 M$_{\odot}$\\
         Maximum $|$S$_z|$ & 0.05\\
         Lower frequency cutoff & 30 Hz\\    
         Approximant & TaylorF2\\
         Minimum match & 0.98\\
         Total templates & 30,858\\
         \hline
         
    \end{tabular}
    \caption{Parameters used to create the template bank, and the total number of templates produced.}
    \label{tab:template_bank}
\end{table}

\subsection{Noise data selection and glitch identification}
\label{sec:noise}

To generate our datasets, we fetched O3 public data in 1 week chunks \citep{gwosc}, and only fetched segments of data where both LIGO Hanford and LIGO Livingston are online, as these were the most sensitive detectors during O3. While a single-detector search pipeline would be useful for when one of the LIGO detectors is down, we currently only consider the two-detector case as this configuration is more sensitive, and handling arbitrary pairs of interferometers adds complexity. We choose to process data in segments of 1,024 seconds, so segments shorter than this duration are excluded. We also exclude segments of data containing GW signals from our training datasets, as identified by the GWTC catalogs \citep{gwtc1,gwtc2,gwtc3}. Neither of these restrictions significantly reduces the amount of data available. For the first week of O3, these criteria lead to a duty factor of $\sim55\%$, which is comparable to the combined duty factor of the two LIGO detectors during O3a. Based on the selected segments of data, we then produce a list of valid integer GPS times with the only requirements being that the merger time is at least 100 seconds from the start of the segment (slightly longer than the longest possible waveform), and at least 24 seconds from the end of the segment. These valid GPS times are used for assigning training and validation samples a merger time. The only preprocessing we apply to the data is downsampling it to 2,048 Hz. We compute the PSD for each week of data, which is used for normalizing the SNR time series. We use the Welch method to compute the PSD with a 4-second window for each 1,024-second segment of data. These PSDs are then averaged together for a mean PSD for each week of data.

We also split the valid GPS times into GPS times where a sample would contain a glitch, and GPS times where a sample would not contain a glitch. We used Omicron to identify the glitches in the noise \citep{Robinet2020}. Glitches with SNR $<$ 6 or a maximum frequency less than 30 Hz were ignored. We save the glitch GPS time, SNR, peak frequency and frequency range for use in our training dataset construction.

\subsection{Training dataset construction}
\label{sec:dataset_construction}

Once the noise for a training set is downloaded and glitches in the noise are identified, the parameters for the waveforms to be injected are sampled. The signal parameter ranges are shown in Table~\ref{tab:parameters}, and the priors were constructed and sampled using the \textsc{Bilby} library \citep{Ashton2019}. We use a distance distribution that is directly proportional to the distance, instead of an astrophysical distance distribution that is uniform in comoving volume. This has the effect of increasing the number of medium to high SNR signals compared to the astrophysical distribution. This distribution was chosen because we found from early tests that our astrophysical training sets had very few high SNR signals, which negatively impacted the neural network's performance on them. We use rejection sampling to repeatedly generate signals until we have reached the target number of signals with an injection network SNR $ > 6$, where network SNR is the quadrature sum of the two detectors' SNRs. This ensures our datasets do not include excessive samples with low SNR.

For our training dataset, we next assign a random GPS time from the first week of O3 to each event. To ensure we have enough unique noise realizations, we use a different random GPS time for each interferometer. We also specify the probability for a sample to contain a glitch. Based on the number of glitches found by Omicron, we chose the Hanford glitch fraction to be 0.1, and the Livingston glitch fraction to be 0.15, as glitches occurred more frequently in Livingston \citep{gwtc3}. These glitch fractions are around an order of magnitude higher than the actual rate of glitches during O3, but we specified these fractions to ensure the neural network is sufficiently trained on glitches. 

To ensure samples intended to contain a glitch have glitch power present in the SNR time series, we offset the GPS time for glitchy samples based on the peak glitch frequency. If a glitch is present exclusively at $f$ Hz, then the glitch will only be present in the SNR time series $t(f)$ seconds before the end of the matched filtering template, where to the first order 
\begin{equation}
    t(f) = \frac{5 c^5}{256 \pi^{8/3} G^{5/3}} \frac{1}{\mathcal{M}_c^{5/3} f^{8/3}}\,,
\end{equation}
\noindent
where $\mathcal{M}_c = (m_1 m_2)^{3/5}/ (m_1 + m_2)^{1/5}$ is the chirp mass of the template. For BBH templates, the response of the glitch at different places in the template is not a significant issue, as BBH mergers are only detectable for $\mathcal{O}$(1) second, but for BNS mergers a low-frequency glitch at the merger time would be present tens of seconds later in the SNR time series, and would not affect the SNR around the signal. 

Once a sample has been assigned intrinsic and extrinsic parameters, a set of templates are selected for matched filtering based on the sample's intrinsic parameters. Since the overlap between a BNS waveform and a template is very sensitive to their chirp mass mismatch, most templates in the template bank will produce a negligible response to an injection waveform. The inclusion of SNR time series from signals filtered with low-overlap templates would be counterproductive for the training set, as they would be functionally indistinguishable from noise. To avoid including these low-overlap samples, we select templates using \textsc{PyCBC}'s \texttt{get\_point\_distance} function, which can approximate the overlap between waveforms given the intrinsic parameters. For every training injection, we sort the point distances from each template to the injection waveform's parameters, then select the template with the lowest point distance. Nine other templates are randomly sampled from the 100 next closest templates, to ensure a spread of overlaps for each injection waveform. We found that with this method the overlap between the signal and the template is at least 0.5 for all injections in our training dataset. 

Strain samples with injections are created by first generating the injection waveform with the SpinTaylorT4 approximant \citep{Buonanno2009}. The waveforms are then projected onto the detectors and placed with the merger at the center of a 1024s-long noise segment based on the Hanford GPS time, with the Livingston detector signal offset based on the extrinsic parameters. The strain is then filtered with the 10 selected templates, and the resulting SNR time series are then sliced down to a 2-second time series centered on the merger. As our neural network's input window is 1 second long, this 2-second slice of SNR time series allows the merger time to be placed randomly within the input window. However, to ensure the entire peak from the merger in both interferometers is completely contained within an SNR time series, we place the merger at least 1/8th of a second from the edges of the window.

For our training dataset, we generate a total of 750,000 SNR time series from 75,000 injection waveforms. We also generate 75,000 random noise samples which are each filtered with 10 random templates, for a total of 750,000 noise samples. For our validation dataset, we use 100,000 injection samples and 100,000 noise samples, but only generate the SNR time series of the closest point distance template (a random template for noise samples), for a total of 200,000 validation samples.

\begin{table}
    \centering
    \begin{tabular}{l @{\hskip 0.5cm}l @{\hskip 0.5cm} l}
         \hline
         \hline
         Parameter& Prior & Range  \\
         \hline
         m$_1$,m$_2$ (M$_{\odot}$)& Uniform & [1, 2.6]\\
         S$_{1z}$, S$_{2z}$ & Uniform & [-0.05, 0.05]\\
         Distance (Mpc) & $d$ & [2, 400]\\
         Right ascension & Uniform & [0, 2$\pi$]\\
         Declination & Cosine & [0, $\pi$]\\
         Inclination & Sine & [0, $\pi$]\\
         Polarization & Uniform & [0, $\pi$] \\
         Network SNR & $\cdots$ & $\geq$ 6 \\
         \hline
    \end{tabular}
    \caption{Parameter ranges and priors used to create the training sets. The component masses are swapped if necessary to ensure m$_1$ is larger. Network SNR is estimated from a sample's parameters, and the sample is only accepted if it meets the threshold.}
    \label{tab:parameters}
\end{table}

\section{The Neural Network}

\label{sec:NN}

\subsection{Model architecture}
\label{sec:model_arch}
\begin{figure*}
    \centering
    \includegraphics[width=0.8\linewidth]{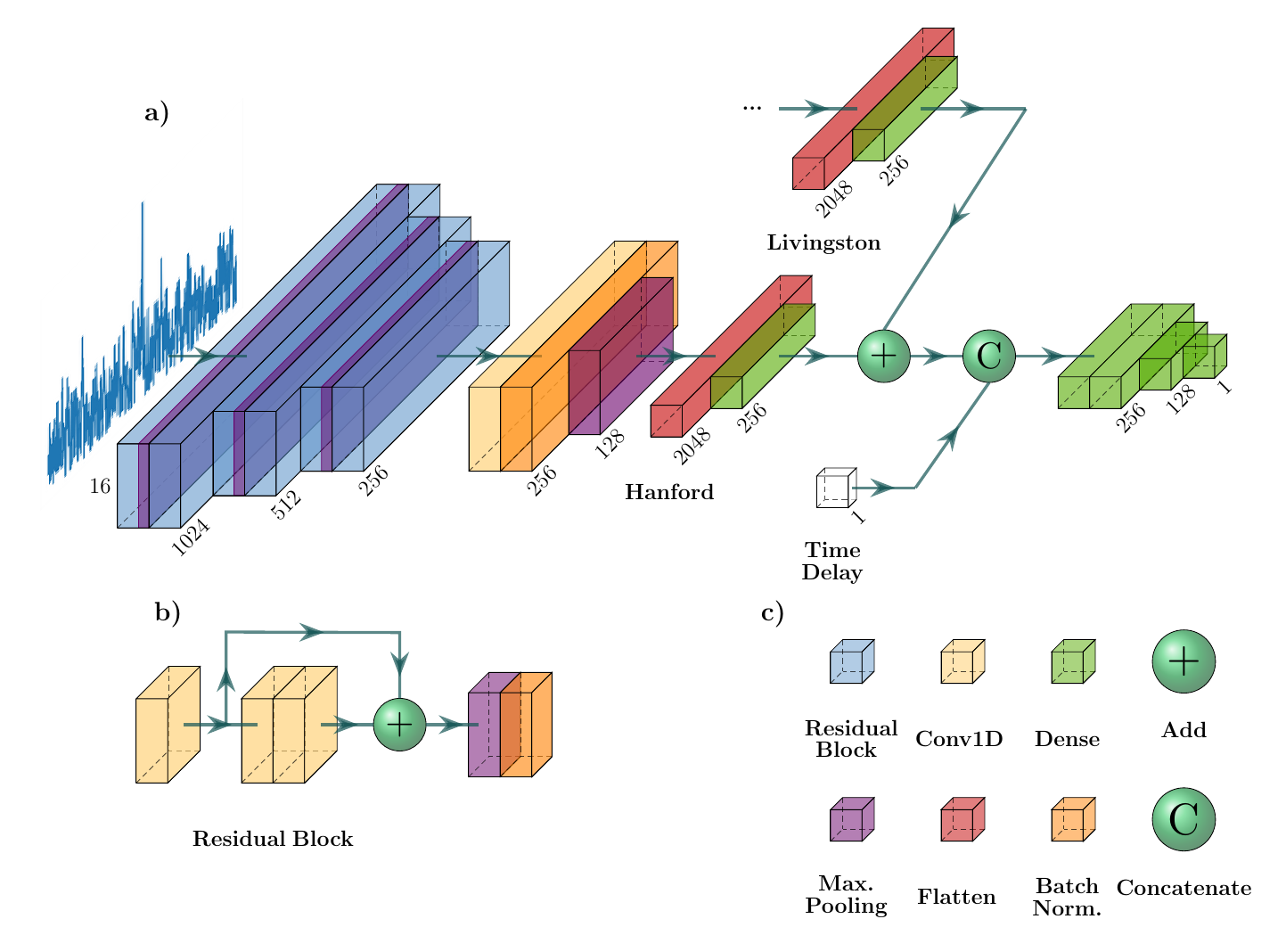}
    \caption{(a) The high-level architecture of our neural network. Each block represents a layer or group of layers, with the height of the block representing the number of convolution kernels and the width representing output size of the layer. The Hanford input branch is shown, and the Livingston input branch has the same architecture. (b) The internal structure of each residual block. (c) A legend of layers and their representation in the diagram. Residual blocks with a purple stripe contain a maximum pooling layer. During training, there is a dropout layer between the dense layers, and a sigmoid activation layer after the last dense layer.}
    \label{fig:model}
\end{figure*}

The two-detector model we use is a convolutional neural network (CNN), composed primarily of residual blocks \citep{He2015}. Its broad structure is two independent branches of residual blocks, one for each interferometer's SNR time series. Each branch takes 1 second of 2,048 Hz SNR time series from its corresponding interferometer. The branches have identical architecture, but have different layer weights from training. These branches are concatenated with an addition layer, which is then followed by four dense layers. The network was built and trained using the Tensorflow library \citep{tensorflow2015-whitepaper}. Figure~\ref{fig:model} shows a high-level representation of the neural network's architecture. 

The network is constructed such that it can be split into three subnetworks around the addition layer between the two branches. The three resulting networks (hereafter called the H, L and combiner subnetworks) can then be operated independently. Since the combiner subnetwork is a simple dense network, it can be run much faster on a CPU than the H and L subnetworks that require a GPU to run efficiently. The advantage of this network architecture is that the outputs of the H and L subnetworks can be independently paired up before being input into the combiner subnetwork, which aids in constructing our background to characterize our ranking statistic, as further explained in Sec.~\ref{sec:bg}. One downside of this architecture is that any coincidence information between the two detectors is lost. To mitigate this, an additional input $\Delta t$ is passed to the combiner subnetwork as well as the H and L subnetwork outputs 

\begin{equation}
    \label{eq:dt}
    \Delta t = \frac{1}{| t_{\text{max}(\text{H1})} - t_{\text{max}(\text{L1})}| + 0.05}\,,
\end{equation}
\noindent
where $t_{\text{max}}$ is the time of the peak SNR of an interferometer. 

During training, there are three additional features that are removed during inference. First, we use a dropout layer between each dense layer in the combiner subnetwork to reduce overfitting on the training set. Second, we use a sigmoid activation layer after the final dense layer to constrain the prediction to between 0 and 1. Removing this activation during inference has been shown to be effective at mitigating the resolution limitations of 32-bit precision \citep{Schafer2022a} and allows our ranking statistic to be unbounded. Thirdly, we add a custom layer before the sigmoid activation layer which divides the output of the last dense layer by a factor of 4, which helps prevent the sigmoid layer from rounding predictions to 0 or 1 during training.

\subsection{Training}

The network was trained using the training and validation datasets described in Sec.~\ref{sec:dataset_construction}. The training and validation samples were weighted in training if their network SNR was greater than 10 by a factor of SNR/10, as we found that the network tended to assign very low prediction values to very high SNR signals, likely because it was mistaking them for glitches. This sample weighting factor also applied to noise-only samples, and so the neural network was additionally penalized for labeling SNR time series with high SNR glitches as signals.

The network was trained with the binary crossentropy loss function using the ADAM optimiser \citep{Kingma2017}, and an initial learning rate of $10^{-4}$. We use the Keras callbacks ReduceLROnPlateau and EarlyStopping to fine-tune the network further: if the validation loss did not improve in 15 epochs, the learning rate was halved, and training stopped if the network's validation loss did not improve after 25 epochs. We created a custom metric, which we call LogAUC, for EarlyStopping to determine when the network should stop being trained. LogAUC is based on Keras's \texttt{AUC} metric, which calculates the true alarm probability and false alarm probability at different thresholds (i.e., a receiver operating characteristic, or ROC curve), then computes the area under the resulting curve. Maximizing the area under this curve ensures a trained model is sensitive at the tested false alarm thresholds. LogAUC simply spaces the false alarm rate thresholds logarithmically and calculates the area of the curve in log space, rather than linearly, which ensures that the sensitivity at low false alarm rates is given suitable weight during training. This is important as GW detection requires a focus on low false alarm rates assigned over many orders of magnitude. Training the network typically takes 3 hours on an NVIDIA A100 GPU.

\section{Search Method and False Alarm Rate Assignment}
\label{sec:bg}

Here we describe how we run our search pipeline and compute our ranking statistic, which is used for assigning false alarm rates to events. Since we detect in the SNR time series rather than the strain, our ranking statistic involves several data reduction steps to reduce the computational requirements of the neural network. The ranking statistic for a second of data is computed with the following workflow. First, the SNR time series are computed using the entire template bank. These SNR time series are computed in clusters of 30 due to memory limitations, and are ordered by chirp mass (i.e., all templates in the cluster have a similar chirp mass). For each second of SNR time series, we use a coincident peak-finding algorithm to find the highest network SNR of the time series. A trigger is produced if a peak is found with SNR $> 4$ in at least one of the detectors. Only one trigger is saved per SNR time series per second.

The SNR time series with the highest network SNR trigger from each cluster is then used as the input to the neural network. The neural network then makes 16 predictions per second on the SNR time series, i.e., an inference rate of 16 Hz. The moving average of each series of predictions is then used as the ranking statistic on that second of data with that template. Using an inference rate greater than 1 Hz aids with separating glitches and high SNR noise from signal candidates, and this benefit has been noted in previous works \citep{Beveridge2024, Marx2024, Koloniari2024}. A noise peak is unlikely to produce multiple high-valued predictions in a row, while a peak from a signal would consistently produce high predictions while in the input window. More information on the effect of the inference rate on sensitivity can be found in Sec.~\ref{sec:IR}. The moving average predictions from each of the template clusters are then compared, and the highest moving average prediction becomes the ranking statistic for that second. In addition to the ranking statistic, the network SNR of the trigger, the time of the peak SNR and the ID of the template that produced the trigger are saved for each second of data.

To characterise our ranking statistic background, we collect the ranking statistic on one week of noise, for which we use the third week of O3. To extend our background, we perform 200 time shifts for each second of data. Time shifts are a simple method for increasing the size of a background, thus making it a more accurate distribution, without requiring an excessive amount of real detector data. Time shifts typically involve shifting the data of two interferometers by a time greater than their light travel time and then computing the ranking statistic of this new pair of data. In our case, however, time shifting the SNR time series would be computationally infeasible, as this would result in a two-hundredfold increase in the number of predictions our neural network would have to make. Our solution to this is to divide the network into the H, L and combiner subnetworks (as described in Sec.~\ref{sec:NN}), and time shift the outputs of the H and L subnetworks when collecting our background. The benefit of splitting the model this way is that only the combiner subnetwork has to process 200 time shifts, and the H and L subnetworks only process the single week of background data. Since the combiner subnetwork is much less computationally expensive than the H and L subnetworks, it can predict on 200 time shifted samples when running on a CPU at the same pace as the H and L models running on a GPU. This procedure is similar to that described in \citep{Schafer2022b}. In our implementation, we keep the Hanford SNR time series fixed and shift the Livingston SNR time series by one second for each time shift.

\begin{figure}
    \centering
    \includegraphics[width=\columnwidth]{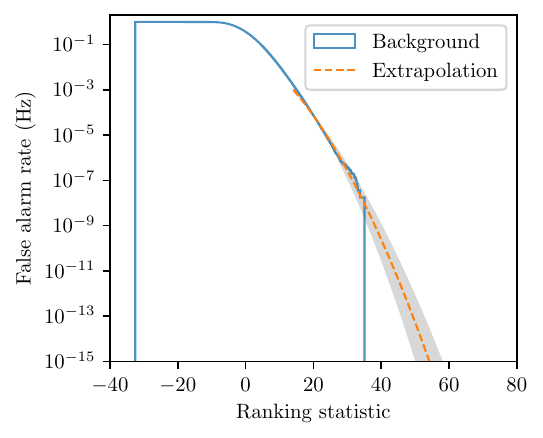}
    \caption{Ranking statistic distribution from 1.8 years of background data, and the corresponding false alarm rates. The orange dashed line shows the Gaussian fit used to assign false alarm rates to ranking statistics higher than the background. The gray region shows the range of the extrapolations assigned using 10 independent subsets of the background.}
    \label{fig:Background}
\end{figure}

Now that we have collected a noise background, we can assign a false alarm rate to new triggers. Since we produce one trigger per second, if a trigger's ranking statistic falls within the background, the false alarm rate of a trigger is simply the fraction of background points with a higher ranking statistic than the trigger. With the week 3 background alone, we can assign false alarm rates down to $1.77 \times 10^{-8}$ Hz, or 0.56/yr. To assign false alarm rates to new triggers that have a higher ranking statistic than the highest background point, we implement an extrapolation of our background. Since we found that our ranking statistic background distribution was approximately Gaussian, we chose to fit a Gaussian extrapolation to the tail of the distribution. We only fit to the tail as ideally the extrapolation should smoothly extend the background, but fits with the whole background introduced a discontinuity between the tail of the background and the extrapolation. The extrapolation was fit to points with a false alarm rate below $10^{-3}$ Hz, and the best fit had an R$^2 = 0.984$. The background and extrapolation used are shown in Fig.~\ref{fig:Background}. While this extrapolation is naturally less accurate than performing a longer background run, it is computationally infeasible to produce backgrounds with our method longer than a few years. To verify that our extrapolation is a reasonable extension of our background, we split our background into 10 independent subsets and fit an extrapolation to each of them. We find that the full background's extrapolation is an accurate marginalization of the subsets, as it lies at the midpoint of the range of the subsets. Note that the detection threshold of 1 per 2 months \citep{gwtc3} used in this work falls within our noise background. Thus, the extrapolation used does not affect whether an event is detected.

\begin{figure}
    \centering
    \includegraphics[width=\columnwidth]{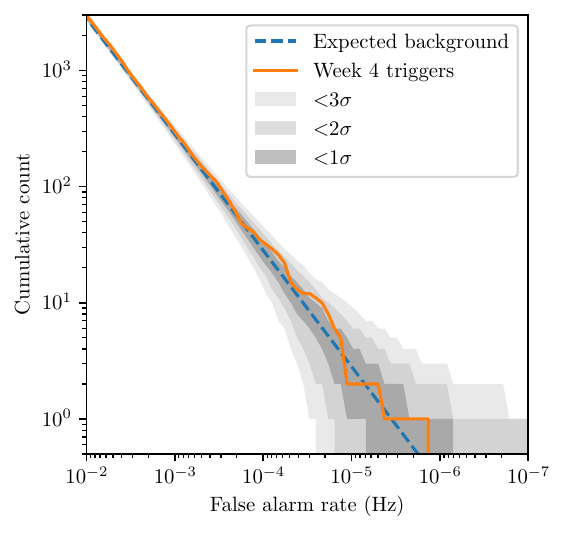}
    \caption{Cumulative counts of background triggers vs. false alarm rate for the fourth week of O3. The false alarm rates were assigned with background collected in the third week. The shaded areas show different levels of Poisson uncertainty.}
    \label{fig:cfar}
\end{figure}

To validate our false alarm rate assignment, we then performed a background run in the fourth week of O3 without time shifting the data. The false alarm rates for the fourth week triggers were assigned using than the third week background, the cumulative counts of which are shown in Fig.~\ref{fig:cfar}. Since the fourth week background is within the $3 \sigma$ Poisson uncertainty bounds of the expected background, we find that our false alarm rate assignment for this week is accurate.

\section{Results}
\label{sec:inj}

\subsection{Injection run results}

We measure our pipeline's performance using an injection run, which is a search for simulated signals placed into noise. Our injection run data is the set of BNS injections from the GWTC-3 catalog that was used to estimate the offline detection pipeline sensitivities in O3 \citep{gwtc3,GWTC3-data}. The parameters of these injections can be found in Table~\ref{tab:O3_parameters}. Since we used the third week of O3 for collecting our background, we selected the fourth week of O3 for our injection run. After applying the same cuts to the data as described in Sec.~\ref{sec:noise}, we were left with a set of 2,800 injections to test our search's sensitivity with. The injection run was performed using the same procedure described in Sec.~\ref{sec:bg}, with the only changes being adding the injections to the noise before matched filtering, and the lack of time shifts. We then assign a false alarm rate to the injections using the collected background, or the Gaussian extrapolation if the ranking statistic is greater than the highest background point. 

\begin{table}
    \centering
    \begin{tabular}{l @{\hskip 0.5cm} l @{\hskip 0.5cm} l}
         \hline
         \hline
         Parameter& Prior & Range  \\
         \hline
         m$_1$ (M$_{\odot}$)& m$_1$ & [1, 2.5]\\
         m$_2$ (M$_{\odot}$)& Uniform & [1, m$_1$]\\
         S$_{1}$ & Isotropic & [-0.4, 0.4]\\
         S$_{2}$ & Isotropic & [-0.4, 0.4]\\
         Redshift & $ (1+z)^{-1}$ & [0, 0.15]\\
         Right ascension & Uniform & [0, 2$\pi$]\\
         Declination & Cosine & [0, $\pi$]\\
         Inclination & Sine & [0, $\pi$]\\
         Polarization & Uniform & [0, $\pi$] \\
         Network SNR & $\cdots$ & $\geq$ 6 \\
         \hline
    \end{tabular}
    \caption{BNS parameter priors and ranges of the GWTC-3 O3 search sensitivity injection dataset.}
    \label{tab:O3_parameters}
\end{table}

\begin{figure}
    \centering
    \includegraphics[width=\columnwidth]{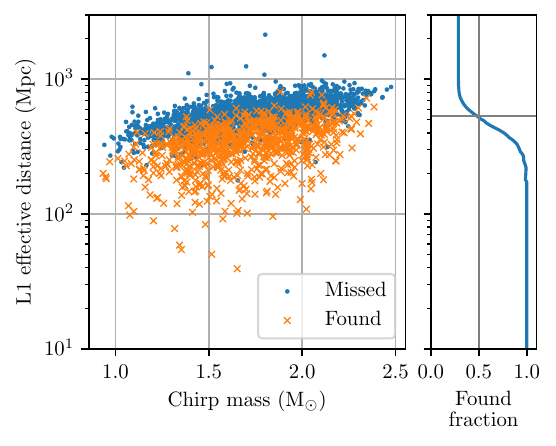}
    \caption{Left: missed and found injections against their detector-frame chirp mass and effective distance in the Livingston detector. An injection is classed as found if its assigned false alarm rate is less than 1 per 2 months. Right: fraction of injections that were found at or below a given Livingston effective distance.}

    \label{fig:deff}
\end{figure}

The next step is to determine when a BNS merger has been detected by the pipeline. We consider an injection detected, or ``found'', if there is a trigger within one second of the injection time below a false alarm rate threshold of 1 per 2 months ($\sim2 \times 10^{-7}$ Hz), the O3 open public alert threshold \citep{gwtc3}. Figure~\ref{fig:deff} shows the distribution of events that were missed or found at this threshold, with respect to their effective distance in the most sensitive detector, Livingston, and detector-frame chirp mass. As expected, our pipeline tends to find closer events, and is able to find events with a higher chirp mass out to further distances, since heavier binaries have a larger GW amplitude than lighter binaries at the same distance. As shown in Fig.~\ref{fig:deff}, the pipeline finds 34\% of all injections in the dataset, 50\% of all events closer than an effective distance of 530 Mpc, and 100\% of all events closer than an effective distance of 175 Mpc. The closest missed event had a recovered Livingston SNR that was 50\% higher than its injected SNR due to a loud blip glitch. This indicates there is a limit to how loud a glitch can be before the pipeline is unable to recover the event.

\begin{figure}
    \centering
    \includegraphics[width=0.85\columnwidth]{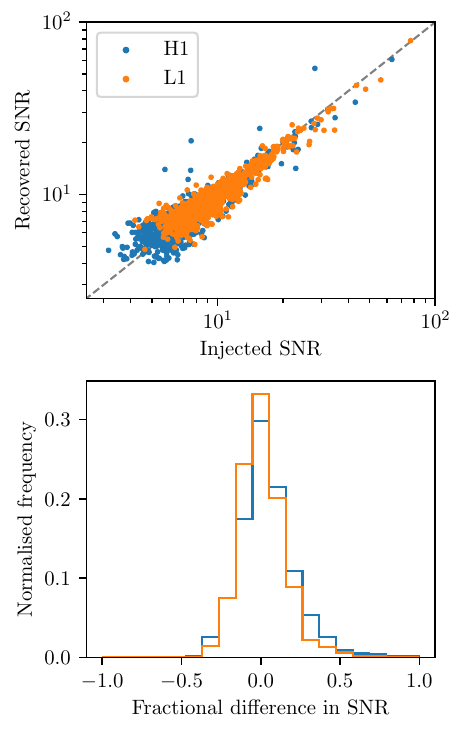}
    \caption{Top: injected SNR of the found events against the recovered SNR for the Hanford (H1) and Livingston (L1) detectors. Bottom: fractional difference between the injected and recovered SNR for each detector.}
    \label{fig:snr_rec}
\end{figure}

Figure \ref{fig:snr_rec} shows the recovered Hanford and Livingston SNR of the detected events. The recovered SNR for each detector is close to the injected SNR, as expected, where the injected SNR is the optimal matched filtering SNR of a signal with the given injection parameters. The spread at low SNRs is due to noise power having a greater influence on the recovered SNR, so individual noise realizations can cause the recovered SNR to noticeably increase or decrease. Six outlier events had a significantly higher recovered Hanford SNR than their injected SNRs. In all of these events, there was a loud blip glitch in the Hanford detector between 1 and 25 seconds before the merger time. This indicates that the neural network is capable of making detections even when a glitch significantly affects the SNR time series. However, the glitchy event missed at $\sim$ 175 Mpc shows that it is possible for loud signals to be missed due to loud glitches.

\begin{figure}
    \centering
    \includegraphics[width=0.85\columnwidth]{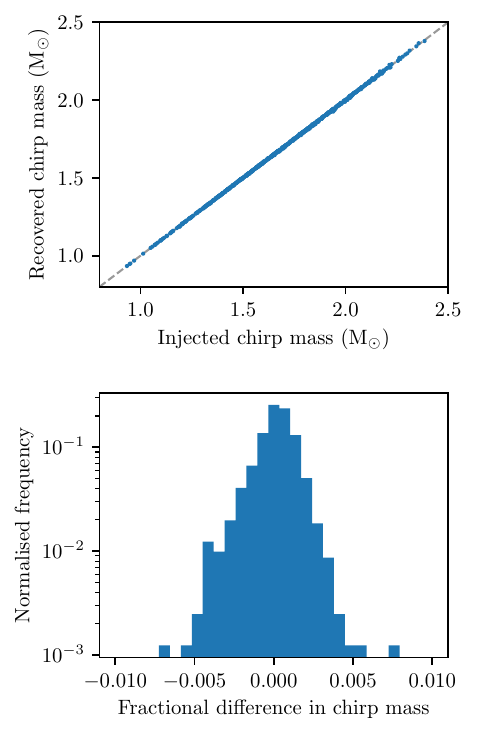}
    \caption{Top: injected detector-frame chirp mass against recovered detector-frame chirp mass for all found events. Bottom: fractional difference in detector-frame chirp mass for the found events. All found events have a recovered chirp mass within 1\% of the injected chirp mass.}
    \label{fig:cm_rec}
\end{figure}

The chirp mass of an event can also be estimated from the parameters of the template that produced the trigger. Figure~\ref{fig:cm_rec} shows the chirp mass of found events closely matches the injected chirp mass. The mean mismatch was $1.23 \times 10^{-4}$, the standard deviation of the mismatch was $1.4 \times 10^{-3}$, and the highest mismatch in recovered chirp mass was 0.7\%, which is similar to the BNS chirp mass recovery of other detection pipelines \citep{Chu2022,Ewing2024}. This shows that our template bank is sufficiently dense to cover all regions of the tested parameter space, and that the ranking statistic is highest on SNR time series from templates with parameters that closely match the true event parameters. Note that these chirp masses are detector-frame rather than source-frame, as gravitational waves are subject to redshift. The actual source-frame chirp mass of a detected event needs to be estimated from distance information.

\subsubsection{Search sensitivity}

Using the false alarm rates assigned to the injections, we now measure the sensitivity of our search. For this, we use the sensitive volume, which measures the volume over which a search method would expect to detect signals at a given false alarm rate. The sensitive volume is defined by

\begin{equation}
\label{eqn:sensitive_vol}
    V(\mathcal{F}) = \int   \epsilon (\mathcal{F}; \textbf{x}, \theta) \phi(\textbf{x}, \theta) \text{d}\textbf{x} \text{d}\theta\,,
\end{equation}
\noindent
at a false alarm rate $\mathcal{F}$ where $\theta$ and $\textbf{x}$ are respectively the physical and spatial parameters of the injections, $\epsilon$ is the recovery fraction of signals, and $\phi$ is the spatial distribution of the injection population \citep{Usman2016}.

In the case of the O3 injection set, the injections are rejected (i.e. treated as missed) if their network SNR is less than 6, and are uniformly distributed in comoving volume. For this set of injections, the sensitive volume is estimated as

\begin{equation}
\label{eqn:sensitive_vol_simplified}
    V(\mathcal{F}) = \frac{1}{N_{\text{draw}}} \sum_{i=1}^{N_{\text{det}}}  \frac{p(\theta_i)}{p_{\text{draw}}(\theta_i, z_i)} \frac{dN}{dz} (z_i)\,, 
\noindent
\end{equation}

where $N_{\text{draw}}$ is the number of drawn injections (rejected injections count toward $N_{\text{draw}}$), $N_{\text{det}}$ is the number of injections with $\mathcal{F}_{i} \leq \mathcal{F}$, $p_{\text{draw}}$ is the draw probability of an injection, and $p(\theta_i)$ is the parameter distribution used \citep{GWTC3-data}. The Monte Carlo uncertainty of the sensitive volume estimate is given by \citep{Farr2019}

\begin{equation}
    \sigma^2_V = \frac{1}{N_{\text{draw}}^2} \sum_{i=1}^{N_{\text{det}}}  \left( \frac{p(\theta_i)}{p_{\text{draw}}(\theta_i, z_i)} \frac{dN}{dz} (z_i) \right)^2 - \frac{\langle V \rangle^2}{N_{\text{draw}}}\,.
\end{equation}
\noindent
In the rest of this work, we also report the sensitive distance for a given sensitive volume, as this quantity is more commonly used for BNS searches \citep{Schafer2020, DalCanton2021}.

\begin{figure}[h]
    \centering
    \includegraphics[width=\columnwidth]{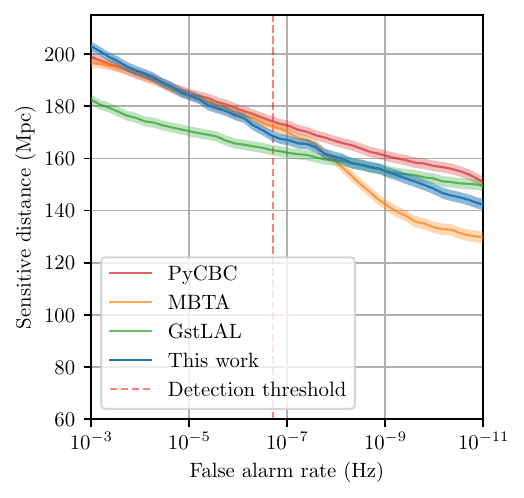}
    \caption{Two-detector sensitive distance for our pipeline compared to the current detection pipelines for the third week of the O3 injection run. The 1 per 2 months detection threshold is marked with a dashed line. Note that the pipeline results are a subset of a larger search, which reduces their sensitivity to BNS mergers. Pipeline data from \citep{GWTC3-data}.}
    \label{fig:dsens}
\end{figure}

To compare our method to the detection pipelines, we take the pipeline's reported false alarm rates for the set of 2,800 injections and calculate their sensitive distances using Eq.~(\ref{eqn:sensitive_vol_simplified}). We only compare to the PyCBC, GstLAL and MBTA pipelines, as the SPIIR pipeline did not run offline in O3, and the unmodeled cWB pipeline is not sensitive to BNS mergers. As we do not consider periods where one or both of the LIGO detectors are offline, the following sensitivities can be interpreted as the sensitive distance of the pipelines and our method when the Hanford and Livingston detectors are observing. One caveat for the pipeline sensitivities is that in O3 the existing pipelines used an additional metric, $p_{\text{astro}}$, to determine the likelihood that a trigger is of astrophysical origin. The sensitivity results reported in \citep{gwtc3} only focus on triggers with $p_{\text{astro}} > 0.5$, but since we do not compute a $p_{\text{astro}}$, we consider the false alarm rates for all of the pipelines irrespective of their estimated $p_{\text{astro}}$ values. Another caveat is that the other search pipelines also search for BBH and NSBH events, which slightly reduces their sensitivity to BNS events (further discussed in Sec.~\ref{sec:direct}).  Figure~\ref{fig:dsens} shows the two-detector sensitive distances of the offline detection pipelines and our method's sensitive distance during the week 4 injection run. We find that our pipeline has a similar sensitivity to the PyCBC and MBTA pipelines above the 1 per 2 months detection threshold, and a similar sensitivity to the PyCBC and GstLAL pipelines below the detection threshold. 

\begin{figure}
    \centering
    \includegraphics[width=\columnwidth]{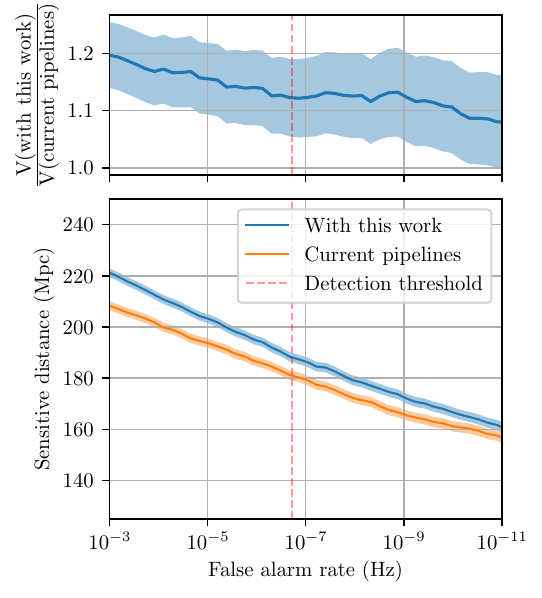}
    \caption{The combined sensitivity of the pipelines with and without our pipeline. ``Current pipelines" includes events that were detected by any of the existing pipelines (PyCBC, MBTA, or GstLAL), and ``with this work" includes events that were detected by any of the existing pipelines or by our pipeline. Top: fractional sensitive volume increase as a function of false alarm rate. Bottom: sensitive distance of both configurations as a function of false alarm rate. The 1 per 2 months detection threshold is marked with a dashed line.}
    \label{fig:total_dsens}
\end{figure}

While the sensitive distance of our method is comparable to the online pipelines' sensitivities at all false alarm rates, this metric does not completely explore the expected BNS sensitivity improvement of adding our pipeline to the set of detection pipelines. Since in a real search an event is considered detected when found by one or more pipelines, and the pipelines do not necessarily detect the same events as each other, a useful quantity is the expected sensitivity increase from adding our pipeline to the set of detection pipelines, which is shown in Fig.~\ref{fig:total_dsens}. We find a 12\% increase in the total number of events detected at the 1 per 2 months detection threshold from adding signals detected by only our pipeline to those detected by the three other pipelines. This increase in sensitivity comes from 124 events that were only detected by our pipeline. This compares favorably to the other pipelines, with PyCBC contributing the most single-pipeline detections (78) before considering our pipeline's detections. Additionally, our pipeline detects several events that were previously only detected by one pipeline, and so increases the number of joint detections by 7\%. 

\subsubsection{Comparison to a PyCBC BNS-only search}
\label{sec:direct}

A caveat for this sensitivity comparison is that the offline pipelines search for BNS, NSBH, and BBH events, while our pipeline is currently only searching for BNS events. Searching over a larger parameter space allows for the detection of more sources, but reduces a search's sensitivity to individual source types due to the greater rate of false alarms. Were our pipeline to also search for NSBH and BBH mergers, its BNS sensitivity would be slightly lower. 

\begin{figure}
    \centering
    \includegraphics[width=\columnwidth]{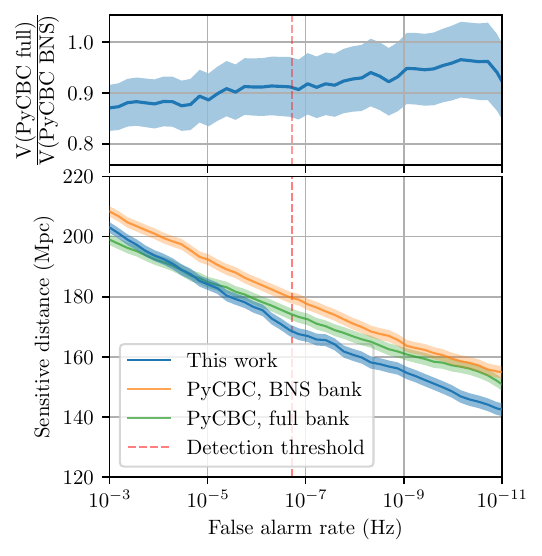}
    \caption{Top: Fractional sensitive volume change between PyCBC with an all source type template bank and our BNS bank. Bottom: The sensitive distance of the PyCBC search with the existing BBH-BNS-NSBH template bank, and with the BNS-only template bank. The sensitive distance of our pipeline is shown for reference.}
    \label{fig:PyCBC_BNS}
\end{figure}

To estimate the sensitivity of our method as if it were part of a larger search for all three source types, we can compare the sensitivity of one of the existing offline searches in two configurations: the existing BBH-BNS-NSBH configuration and a BNS-only search. For this, we ran the PyCBC offline search with our generated BNS template bank on the same injection set in the fourth week of O3. As shown in Fig.~\ref{fig:PyCBC_BNS}, when run with our BNS template bank the PyCBC search is more sensitive, as noise triggers from the other areas of the search space do not decrease its BNS sensitivity. The change in search space from BNS-only to all three source types results in a $\sim 10$\% sensitive volume drop at the detection threshold, with the sensitive volumes converging at lower false alarm rates. From this, we estimate that our BNS search could lose a small amount of its sensitive volume ($\sim$ 10\% at the detection threshold) when scaled up to a larger search for all three source types. 

We also find that our search is less sensitive than PyCBC in the same search space, reaching 83\% of the sensitive volume at the detection threshold. However, by adjusting our pipeline's assigned false alarm rates with a trials factor estimated from the sensitive volume difference of the two PyCBC configurations, we can estimate the total BNS sensitivity increase from adding our pipeline to the set of detection pipelines, as if our pipeline were searching for all three source types. We find that our pipeline would still detect 95 additional BNS events that none of the other pipelines detected, yielding a 9.5\% increase in the total number of BNS events detected. Similarly, it would still increase the number of joint detections by 5.8\%.

We therefore find that while our search and the existing pipelines are searching over different search spaces, this difference does not significantly affect the comparison of our sensitivities. Extending our method to search for all three source types would be a worthwhile future investigation, and would provide an accurate one-to-one comparison to the existing pipelines.

\subsection{Performance on real events}

Here we report on our pipeline's performance for the two confirmed BNS events, GW170817 and GW190425. Detecting these events is a useful test to confirm our pipeline can make the same real event detections as the other search pipelines. 

\subsubsection{GW170817}

GW170817 is an important test of our pipeline for two reasons. First, since GW170817 is in O2 but our neural network was trained on O3 data, it allows us to test whether our pipeline is capable of making detections despite the changes in detector PSD between observing runs. Second, GW170817 is notable for having a loud blip glitch present $\sim$1 second before the merger in the Livingston detector. Searching for GW170817 without removing the blip glitch beforehand will demonstrate whether our pipeline is capable of making real detections without glitch mitigation.

\begin{figure}
    \centering
    \includegraphics[width=\linewidth]{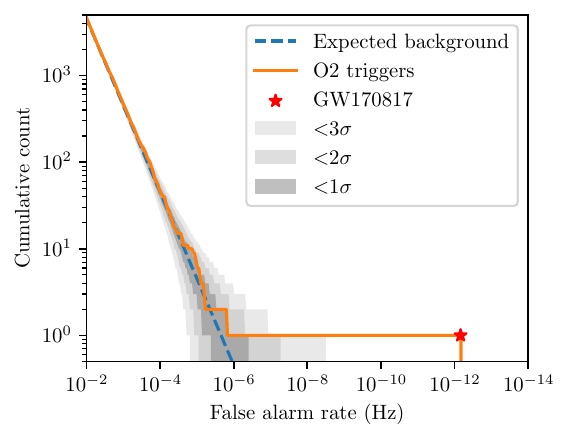}
    \caption{Cumulative count of triggers in the O2 search, against false alarm rate. The triggers are from 1 week of data containing GW170817, and the false alarm rates were assigned by collecting a background of our ranking statistic over the previous week of data.}
    \label{fig:170817_O2}
\end{figure}

\begin{figure}[!]
    \centering
    \includegraphics[width=0.9\columnwidth]{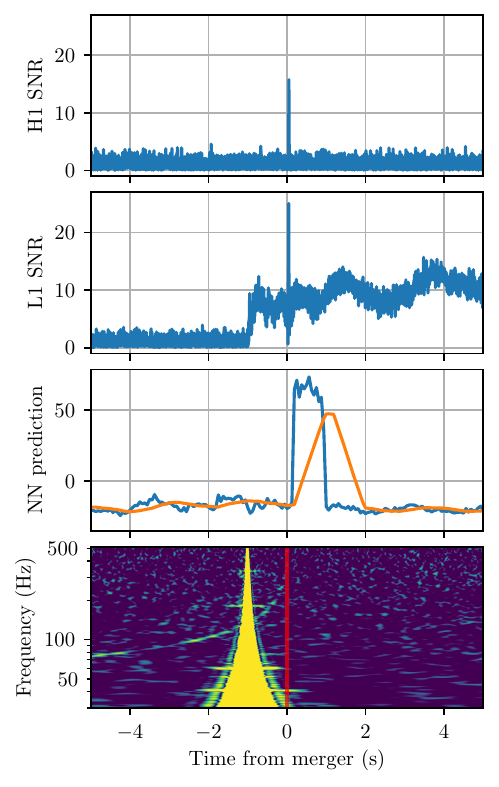}
    \caption{SNR and prediction time series for GW170817. The top two panels show the Hanford and Livingston SNR time series from the triggering template. The third panel shows the 16 Hz neural network prediction time series for the triggering template in blue, and the moving average in orange. For illustrative purposes, the bottom panel shows the time-frequency spectrogram of the Livingston interferometer, with the red line marking the merger time. The loud glitch in Livingston is visible $\sim$1 second before the merger in the spectrogram and the Livingston SNR time series, but did not affect the neural network prediction time series.}
    \label{fig:170817}
\end{figure}

To test our pipeline's ability to detect GW170817, we first performed a 1 week background run in O2, ending 512 seconds before the merger time of GW170817. We then performed a 1 week search immediately after the background run, such that GW170817 was present 512 seconds into the search. Triggers from this search were then assigned false alarm rates with the O2 background run, which are shown in Fig.~\ref{fig:170817_O2}. With this O2 background, GW170817 was assigned a false alarm rate of $6.9 \times 10^{-13}$ Hz, or 1 per 46,000 years. Using the method for estimating the uncertainty of our FAR extrapolation described in Sec.~\ref{sec:bg}, we find that GW170817's FAR is between $1.7 \times 10^{-12}$ Hz and $2.4 \times 10^{-14}$ Hz. GW170817 would be considered confidently detected at any FAR in this range. The triggering template recovered a network SNR of 31.3 and has a detector frame chirp mass of 1.1978 \msun, which is consistent with the previously reported detector frame chirp mass of $1.1977^{+0.0008}_{-0.0003}$ \msun \citep{170817}. The recovered SNR and assigned false alarm rate are also similar to those of the detection pipelines, as shown in Table~\ref{tab:170817_stats}. We also find that the neural network's predictions are insensitive to the glitch present in the Livingston detector, as shown in Fig.~\ref{fig:170817}. The neural network's predictions increase when GW170817 enters the input window, but do not change when the glitch enters the input window. From this, we find that our pipeline is confidently detecting GW170817, and is able to detect it despite the loud glitch.

\begin{table}[h]
    \centering
    \begin{tabular}{l @{\hskip 0.5cm} l @{\hskip 0.5cm} l}
         \hline
         \hline
         
         Search & Network SNR & FAR (Hz)  \\
         \hline 
         
         This work & 31.3 & $ 6.9 \times 10^{-13} $ \\
         PyCBC & 30.9 & $ < 4.0 \times 10^{-13} $ \\ 
         GstLAL & 33.0 & $ < 3.2 \times 10^{-15} $ \\         
         
         \hline
         
    \end{tabular}
    \caption{Search results for GW170817. The other pipeline's reported FARs and SNRs are from \citep{gwtc1}, and were calculated after the Livingston glitch was removed.} 
    \label{tab:170817_stats}
\end{table}

\subsubsection{GW190425}

While GW170817 is relatively straightforward to detect with our method despite the glitch, GW190425 is more challenging. GW190425 was detected as a single detector event by the GstLAL pipeline, as the Hanford detector was offline and the Virgo SNR was below the detection threshold \citep{190425}. Our neural network is only trained on two detector events, and while the Virgo detector was online, we trained the neural network specifically on the two LIGO detectors. However, since we use an addition layer to merge the H and L branches of the neural network (see Fig.~\ref{fig:model}), by setting the output of the H branch to zeroes, the only contribution to the combiner subnetwork will be from the L branch, thus approximating a single-detector pipeline. While we did not train our neural network on the single-detector case, and analysing events in this way was not considered when constructing our method, it provides a compelling proof-of-concept for further investigation into a single-detector extension of our pipeline.

\begin{figure}[!]
    \centering
    \includegraphics[width=0.9\columnwidth]{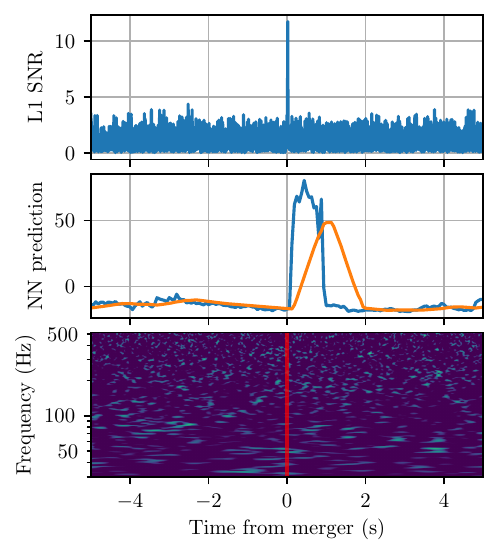}
    \caption{SNR and prediction time series for GW190425. The top panel shows the Livingston SNR time series from the triggering template. The middle panel shows the 16 Hz neural network prediction in blue, and the moving average in orange. The bottom panel shows the time-frequency spectrogram of GW190425, with the red line marking the merger time.}
    \label{fig:190425}
\end{figure}

\begin{figure}
    \centering
    \includegraphics[width=0.85\linewidth]{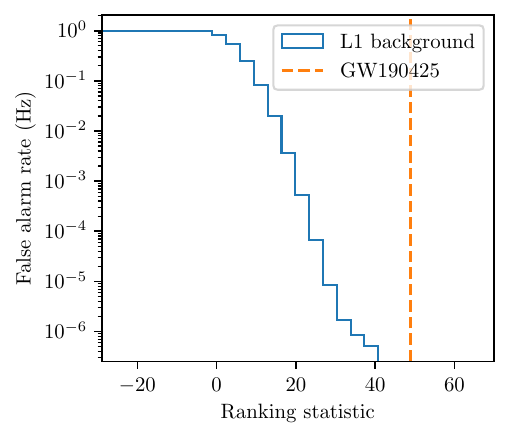}
    \caption{Single-detector ranking statistic background and corresponding false alarm rates for 2.2 months of Livingston noise. GW190425's ranking statistic is shown as the orange vertical line.}
    \label{fig:190425_background}
\end{figure}

To detect GW190425, we first collected a ranking statistic background. Since this is only a single detector event, we cannot use time shifts to extend our background. We therefore collect a background that is longer than 2 months, as 1 per 2 months is our detection threshold. Since the first month of O3 was used for training the model and also contains GW190425, we collected a background in the second, third and fourth months of O3, which yields a 2.2 month-long background. We then ran a search on the fourth week of O3, the week containing GW190425. GW190425 was recovered with an SNR of 11.6, and the triggering template has a chirp mass of 1.488 \msun, which is consistent with GstLAL's recovered chirp mass of 1.487 \msun \citep{190425}. The ranking statistic for the trigger is 48.5, and the SNR and prediction time series for this template are shown in Fig.~\ref{fig:190425}. This ranking statistic is higher than all of the background events, as shown in Fig.~\ref{fig:190425_background}. Since this is only a single-detector event, we conservatively assign a false alarm rate of $ < 1.7 \times 10^{-7} $ Hz (1 per 2.2 months, the length of the background). Table~\ref{tab:190425_stats} shows a comparison of our detection to GstLAL's, the only search pipeline that detected GW190425. We report a similar SNR to GstLAL, but our false alarm rate is several orders of magnitude higher since we do not extrapolate our single-detector background.

\begin{table}[h]
    \centering
    \begin{tabular}{l @{\hskip 0.5cm} l @{\hskip 0.5cm} l}
         \hline
         \hline
         
         Search & Network SNR & FAR (Hz)  \\
         \hline 
         
         This work & 11.6 & $ < 1.7 \times 10^{-7} $ \\
         GstLAL & 12.9 & $ 4.5 \times 10^{-13} $ \\         
         
         \hline
         
    \end{tabular}
    \caption{Search results for GW190425. GstLAL was the only pipeline that detected GW190425, and its reported FAR and SNR are from \citep{190425}.}
    \label{tab:190425_stats}
\end{table}

\subsection{Sensitivity over O3}

\begin{figure*}
    \centering
    \includegraphics[width=0.8\linewidth]{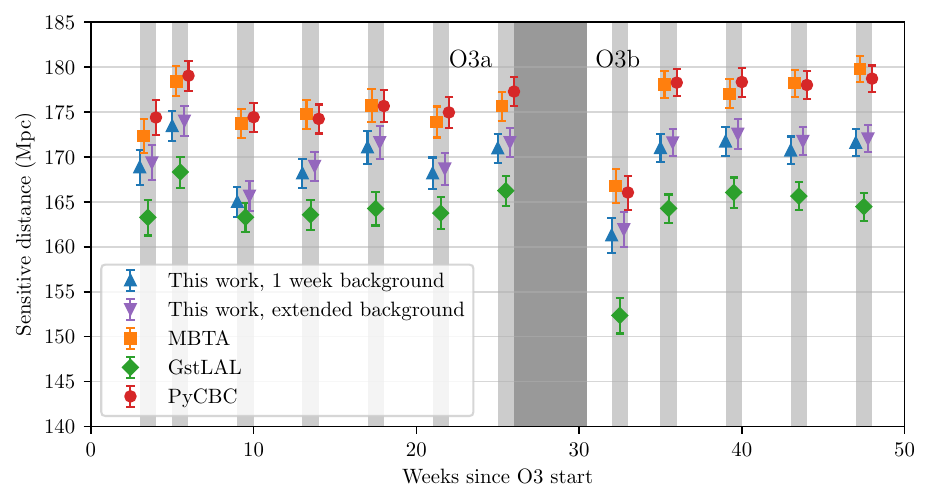}
    \caption{Sensitive distance of our pipeline compared to the offline detection pipelines at the detection FAR threshold of 1 per 2 months over the course of O3. The dark gray region is the break between O3a and O3b, during which the detectors were offline. The light gray regions are the weeks in which the injection runs were performed with our pipeline.}
    \label{fig:sensitivity_over_time}
\end{figure*}

During online and offline searches, the CBC detection pipelines update their noise background over time to compensate for the nonstationarity of the interferometer noise. Since our neural network was trained on the first week of O3 and the background was collected in the third week of O3, we evaluate if there is any noticeable sensitivity loss throughout O3 to determine if the background needs updating or if the network needs retraining on long timescales. Since it would be difficult to determine if any changes in the pipeline's sensitivity week-to-week are due to changes in interferometer sensitivity or due to the neural network performing poorly on noise it was not trained on, we compare the pipeline to the other offline pipelines' sensitivities over O3. For this comparison, we perform additional week-long injection runs using the O3 injection set at a cadence of roughly one per month. Figure~\ref{fig:sensitivity_over_time} shows the sensitive distance of the pipelines over the O3 dataset. Any sensitivity changes in our pipeline are also evident in the other detection pipelines, showing that our pipeline's sensitivity is largely unchanged over the course of O3. 

We also test if our pipeline would benefit from a background collected from multiple times during O3. We perform additional background runs in weeks 16, 32 and 38 of O3, and combine these backgrounds with the week 3 background. When assigning false alarm rates using this extended background instead of just the week 3 background, we find that our sensitivity at the detection threshold is increased, but by less than 1 Mpc in all weeks. This shows that while there is a minor improvement from updating the background with subsequent weeks, the pipeline is capable of running on an entire observing run with a background from a single week of detector data. We also infer that the neural network does not need retraining on unseen noise over long timescales, as we would expect the neural network's sensitivity to decrease over O3 if it did, and that updating the background would have little effect on this sensitivity loss. Despite this, in a real observing scenario, it would still be good practice to continually update the background in case of any substantial shifts in the PSDs of the interferometers.

\subsection{Inference rate comparison}
\label{sec:IR}

As mentioned in Sec.~\ref{sec:bg}, the detection statistic we use is the moving average of 16 neural network predictions, sampled at 16 Hz. In other words, for each second of data, the neural network will make 16 predictions in a sliding window approach. By taking the moving average of multiple predictions on the same second of data, we reduce the impact of spurious high predictions while still allowing multiple high-valued predictions to accumulate from a merger \citep{Marx2024}.

\begin{figure}[!t]
    \centering
    \includegraphics[width=\linewidth]{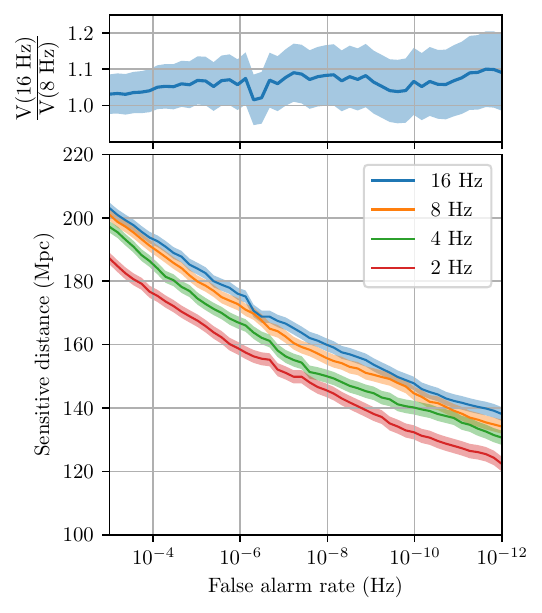}
    \caption{Sensitivity of the pipeline at different inference rates against false alarm rate. Top: fractional sensitive volume increase between a 16 Hz and 8 Hz inference rate. Bottom: sensitive distances at the tested inference rates.}
    \label{fig:inference_rates}
\end{figure}

Here, we compare the sensitivity of the pipeline at different inference rates. Since the computational cost of the background and injection runs are proportional to the inference rate, using a higher inference rate is only beneficial if the sensitivity of the pipeline also appreciably increases. We compared four different inference rates: 2 Hz, 4 Hz, 8 Hz and 16 Hz. Each inference rate was tested independently with the week 4 injection run, the results of which are shown in Fig.~\ref{fig:inference_rates}. The injection run results show that increasing the inference rate improves the sensitive distance at all false alarm rates, but with diminishing returns. While we use a 16 Hz inference rate for all the other results in this work, the $\sim$6\% sensitive volume increase compared to 8 Hz could reasonably be sacrificed to halve the computational requirements of running the pipeline, especially if this search method were adapted for low-latency online detection. 16 Hz was the final inference rate we tested as the doubling of computational resources required for a 32 Hz inference rate would have been unjustified given the minor increase in sensitivity between an 8 Hz and 16 Hz inference rate.

\subsection{Inference speed}

In addition to detecting events in archival data, it is also important to consider the suitability of our method for application to a low-latency online setting. Two important factors in this context are the resource usage required for real-time operation, and any unavoidable latencies associated with the method. We tested the inference speed of our neural network on an NVIDIA A100 GPU on the OzStar supercomputer, and found that it is capable of 34,478 $\pm$ 261 inferences per second. With the template bank, inference rate and clustering method used in this work, our neural network must make 16,457 inferences per second. Therefore, our current implementation would be able to run in low-latency with a single NVIDIA A100 GPU, and the inference step would create $\sim$0.5 seconds of latency. The other unique latency of our method comes from the moving average step. Since the moving average is computed over 1 second of data, it introduces 1 second of unavoidable latency. Aside from these latencies, any other latencies would be dependent on the search pipeline implementation and cannot currently be estimated. Given that our method does not introduce any excessive latencies, we conclude that it is worth investigating its performance in an online implementation.

\section{Conclusions}
\label{sec:discussion}

In this work, we present the most sensitive deep learning-based BNS detection pipeline to date, that is capable of matching the BNS search sensitivities of the offline CBC detection pipelines below the 1 per 2 months detection threshold. When tested with the O3 offline injection set, our pipeline is capable of accurately recovering the SNR and detector-frame chirp mass of injected BNS events. When compared to the other offline pipelines which cover the full stellar-mass binary search space, we find that our pipeline is capable of increasing the total number of BNS detections by 12\%, and increases the number of joint detections by 7\%. We find that our pipeline is capable of detecting both of the real BNS mergers, GW170817 and GW190425, and does not spuriously trigger on the glitch present in GW170817's inspiral. When false alarm rates are assigned using a single week of background, we find that the pipeline's sensitivity does not decrease over the course of O3. From this, we conclude that the pipeline is insensitive to long-term PSD changes over an observing run.

Since our neural network uses SNR time series as its input, it could be implemented into an existing online matched filtering pipeline with relative ease. While we cannot directly compare the latency of our method to the online low-latency pipelines, we find that our method can operate in real-time on a single NVIDIA A100, and would only add 1 second of unavoidable latency after matched filtering. Based on this, we conclude that our method would be a compelling target for implementation in an online pipeline.

One important feature of the detection pipelines that we did not investigate in this work is the ability to detect mergers using an arbitrary set of interferometers. Our current pipeline requires that both Hanford and Livingston are online, but does not take into account Virgo detector data, or periods where one of the LIGO detectors is offline. Since the pipeline was able to detect GW190425 with input only on the Livingston branch of the neural network, training separate combiner models for each combination of active interferometers could allow for the scaling of this method to future observing scenarios. 

Motivated by the results of this study, we aim to apply this method to NSBH mergers in the future. Detecting NSBH mergers with this method would not require any significant changes, apart from the template bank generation method used, as geometric banks are not capable of efficiently covering higher mass regions. A combined model capable of detecting BBH \citep{Beveridge2024}, BNS and NSBH mergers could then be investigated, as this would bring our detection method more in line with the current CBC pipelines which detect all three classes of events, and would improve the accuracy of our comparison to the existing pipelines. We will also investigate the pre-merger detection of BNS and NSBH mergers with this method, with the aim of contributing to early warning triggers to aid multimessenger astronomy with gravitational waves.

\section{Data and Software Availability}

The code we developed for our sample generation is available at \citep{GWSamplegen}. The code we developed for collecting our background and running our search is available at \citep{Infernus}. These repositories can be used to reproduce the results presented in this work.

\section{Acknowledgments}

A.M acknowledges the support of an Australian Government Research Training Program Scholarship while at the University of Western Australia.

The authors thank Ethan Marx for advice on downloading and processing GWOSC data. The authors also thank Ryan Magee and Thomas Dent for their helpful comments and discussions on earlier versions of this manuscript. We also thank Thomas Dent for advice on running the PyCBC search with our template bank.

This work was performed on the OzSTAR national facility at Swinburne University of Technology. The OzSTAR program receives funding in part from the Astronomy National Collaborative Research Infrastructure Strategy (NCRIS) allocation provided by the Australian Government, and from the Victorian Higher Education State Investment Fund (VHESIF) provided by the Victorian Government.

The authors are grateful for computational resources provided by the LIGO Laboratory and supported by National Science Foundation Grants PHY-0757058 and PHY-0823459. 

This research has made use of data or software obtained from the Gravitational Wave Open Science Center \citep{GWOSC_site}, a service of the LIGO Scientific Collaboration, the Virgo Collaboration, and KAGRA. This material is based upon work supported by NSF's LIGO Laboratory which is a major facility fully funded by the National Science Foundation, as well as the Science and Technology Facilities Council (STFC) of the United Kingdom, the Max-Planck-Society (MPS), and the State of Niedersachsen/Germany for support of the construction of Advanced LIGO and construction and operation of the GEO600 detector. Additional support for Advanced LIGO was provided by the Australian Research Council. Virgo is funded, through the European Gravitational Observatory (EGO), by the French Centre National de Recherche Scientifique (CNRS), the Italian Istituto Nazionale di Fisica Nucleare (INFN) and the Dutch Nikhef, with contributions by institutions from Belgium, Germany, Greece, Hungary, Ireland, Japan, Monaco, Poland, Portugal, Spain. KAGRA is supported by Ministry of Education, Culture, Sports, Science and Technology (MEXT), Japan Society for the Promotion of Science (JSPS) in Japan; National Research Foundation (NRF) and Ministry of Science and ICT (MSIT) in Korea; Academia Sinica (AS) and National Science and Technology Council (NSTC) in Taiwan.

\bibliography{bibliography}

\begin{thebibliography}{63}%
\makeatletter
\providecommand \@ifxundefined [1]{%
 \@ifx{#1\undefined}
}%
\providecommand \@ifnum [1]{%
 \ifnum #1\expandafter \@firstoftwo
 \else \expandafter \@secondoftwo
 \fi
}%
\providecommand \@ifx [1]{%
 \ifx #1\expandafter \@firstoftwo
 \else \expandafter \@secondoftwo
 \fi
}%
\providecommand \natexlab [1]{#1}%
\providecommand \enquote  [1]{``#1''}%
\providecommand \bibnamefont  [1]{#1}%
\providecommand \bibfnamefont [1]{#1}%
\providecommand \citenamefont [1]{#1}%
\providecommand \href@noop [0]{\@secondoftwo}%
\providecommand \href [0]{\begingroup \@sanitize@url \@href}%
\providecommand \@href[1]{\@@startlink{#1}\@@href}%
\providecommand \@@href[1]{\endgroup#1\@@endlink}%
\providecommand \@sanitize@url [0]{\catcode `\\12\catcode `\$12\catcode `\&12\catcode `\#12\catcode `\^12\catcode `\_12\catcode `\%12\relax}%
\providecommand \@@startlink[1]{}%
\providecommand \@@endlink[0]{}%
\providecommand \url  [0]{\begingroup\@sanitize@url \@url }%
\providecommand \@url [1]{\endgroup\@href {#1}{\urlprefix }}%
\providecommand \urlprefix  [0]{URL }%
\providecommand \Eprint [0]{\href }%
\providecommand \doibase [0]{https://doi.org/}%
\providecommand \selectlanguage [0]{\@gobble}%
\providecommand \bibinfo  [0]{\@secondoftwo}%
\providecommand \bibfield  [0]{\@secondoftwo}%
\providecommand \translation [1]{[#1]}%
\providecommand \BibitemOpen [0]{}%
\providecommand \bibitemStop [0]{}%
\providecommand \bibitemNoStop [0]{.\EOS\space}%
\providecommand \EOS [0]{\spacefactor3000\relax}%
\providecommand \BibitemShut  [1]{\csname bibitem#1\endcsname}%
\let\auto@bib@innerbib\@empty
\bibitem [{\citenamefont {{Aasi}}\ \emph {et~al.}(2015)\citenamefont {{Aasi}} \emph {et~al.}}]{ALIGO}%
  \BibitemOpen
  \bibfield  {author} {\bibinfo {author} {\bibfnamefont {J.}~\bibnamefont {{Aasi}}} \emph {et~al.} (\bibinfo {collaboration} {LIGO Scientific Collaboration}),\ }\bibfield  {title} {\bibinfo {title} {{Advanced LIGO}},\ }\href {https://doi.org/10.1088/0264-9381/32/7/074001} {\bibfield  {journal} {\bibinfo  {journal} {Classical and Quantum Gravity}\ }\textbf {\bibinfo {volume} {32}},\ \bibinfo {eid} {074001} (\bibinfo {year} {2015})},\ \Eprint {https://arxiv.org/abs/1411.4547} {arXiv:1411.4547 [gr-qc]} \BibitemShut {NoStop}%
\bibitem [{\citenamefont {{Acernese}}\ \emph {et~al.}(2015)\citenamefont {{Acernese}} \emph {et~al.}}]{AVIRGO}%
  \BibitemOpen
  \bibfield  {author} {\bibinfo {author} {\bibfnamefont {F.}~\bibnamefont {{Acernese}}} \emph {et~al.},\ }\bibfield  {title} {\bibinfo {title} {{Advanced Virgo: a second-generation interferometric gravitational wave detector}},\ }\href {https://doi.org/10.1088/0264-9381/32/2/024001} {\bibfield  {journal} {\bibinfo  {journal} {Classical and Quantum Gravity}\ }\textbf {\bibinfo {volume} {32}},\ \bibinfo {eid} {024001} (\bibinfo {year} {2015})},\ \Eprint {https://arxiv.org/abs/1408.3978} {arXiv:1408.3978 [gr-qc]} \BibitemShut {NoStop}%
\bibitem [{\citenamefont {Abbott}\ \emph {et~al.}(2019)\citenamefont {Abbott} \emph {et~al.}}]{gwtc1}%
  \BibitemOpen
  \bibfield  {author} {\bibinfo {author} {\bibfnamefont {B.~P.}\ \bibnamefont {Abbott}} \emph {et~al.} (\bibinfo {collaboration} {LIGO Scientific Collaboration and Virgo Collaboration}),\ }\bibfield  {title} {\bibinfo {title} {{GWTC-1: A Gravitational-Wave Transient Catalog of Compact Binary Mergers Observed by LIGO and Virgo during the First and Second Observing Runs}},\ }\href {https://doi.org/10.1103/PhysRevX.9.031040} {\bibfield  {journal} {\bibinfo  {journal} {Phys. Rev. X}\ }\textbf {\bibinfo {volume} {9}},\ \bibinfo {pages} {031040} (\bibinfo {year} {2019})}\BibitemShut {NoStop}%
\bibitem [{\citenamefont {Abbott}\ \emph {et~al.}(2021)\citenamefont {Abbott} \emph {et~al.}}]{gwtc2}%
  \BibitemOpen
  \bibfield  {author} {\bibinfo {author} {\bibfnamefont {R.}~\bibnamefont {Abbott}} \emph {et~al.} (\bibinfo {collaboration} {LIGO Scientific Collaboration and Virgo Collaboration}),\ }\bibfield  {title} {\bibinfo {title} {{GWTC-2: Compact Binary Coalescences Observed by LIGO and Virgo During the First Half of the Third Observing Run}},\ }\href {https://doi.org/10.1103/PhysRevX.11.021053} {\bibfield  {journal} {\bibinfo  {journal} {Phys. Rev. X}\ }\textbf {\bibinfo {volume} {11}},\ \bibinfo {pages} {021053} (\bibinfo {year} {2021})},\ \Eprint {https://arxiv.org/abs/2010.14527} {arXiv:2010.14527 [gr-qc]} \BibitemShut {NoStop}%
\bibitem [{\citenamefont {Abbott}\ \emph {et~al.}(2023{\natexlab{a}})\citenamefont {Abbott} \emph {et~al.}}]{gwtc3}%
  \BibitemOpen
  \bibfield  {author} {\bibinfo {author} {\bibfnamefont {B.~P.}\ \bibnamefont {Abbott}} \emph {et~al.} (\bibinfo {collaboration} {LIGO Scientific Collaboration, Virgo Collaboration, and KAGRA Collaboration}),\ }\bibfield  {title} {\bibinfo {title} {{GWTC-3: Compact Binary Coalescences Observed by LIGO and Virgo during the Second Part of the Third Observing Run}},\ }\href {https://doi.org/10.1103/PhysRevX.13.041039} {\bibfield  {journal} {\bibinfo  {journal} {Physical Review X}\ }\textbf {\bibinfo {volume} {13}},\ \bibinfo {eid} {041039} (\bibinfo {year} {2023}{\natexlab{a}})},\ \Eprint {https://arxiv.org/abs/2111.03606} {arXiv:2111.03606 [gr-qc]} \BibitemShut {NoStop}%
\bibitem [{\citenamefont {Abbott}\ \emph {et~al.}(2017)\citenamefont {Abbott} \emph {et~al.}}]{170817}%
  \BibitemOpen
  \bibfield  {author} {\bibinfo {author} {\bibfnamefont {B.}~\bibnamefont {Abbott}} \emph {et~al.} (\bibinfo {collaboration} {LIGO Scientific Collaboration and Virgo Collaboration}),\ }\bibfield  {title} {\bibinfo {title} {{GW170817: Observation of Gravitational Waves from a Binary Neutron Star Inspiral}},\ }\href {https://doi.org/10.1103/PhysRevLett.119.161101} {\bibfield  {journal} {\bibinfo  {journal} {Phys. Rev. Lett.}\ }\textbf {\bibinfo {volume} {119}},\ \bibinfo {pages} {161101} (\bibinfo {year} {2017})},\ \Eprint {https://arxiv.org/abs/1710.05832} {arXiv:1710.05832 [gr-qc]} \BibitemShut {NoStop}%
\bibitem [{\citenamefont {{Abbott}}\ \emph {et~al.}(2020)\citenamefont {{Abbott}} \emph {et~al.}}]{190425}%
  \BibitemOpen
  \bibfield  {author} {\bibinfo {author} {\bibfnamefont {B.~P.}\ \bibnamefont {{Abbott}}} \emph {et~al.},\ }\bibfield  {title} {\bibinfo {title} {{GW190425: Observation of a Compact Binary Coalescence with Total Mass {\ensuremath{\sim}} 3.4 M$_{{\ensuremath{\odot}}}$}},\ }\href {https://doi.org/10.3847/2041-8213/ab75f5} {\bibfield  {journal} {\bibinfo  {journal} {Astrophys. J. Lett.}\ }\textbf {\bibinfo {volume} {892}},\ \bibinfo {eid} {L3} (\bibinfo {year} {2020})},\ \Eprint {https://arxiv.org/abs/2001.01761} {arXiv:2001.01761 [astro-ph.HE]} \BibitemShut {NoStop}%
\bibitem [{\citenamefont {{LIGO Scientific Collaboration}}\ \emph {et~al.}(2017)\citenamefont {{LIGO Scientific Collaboration}} \emph {et~al.}}]{170817_multimessenger}%
  \BibitemOpen
  \bibfield  {author} {\bibinfo {author} {\bibnamefont {{LIGO Scientific Collaboration}}} \emph {et~al.},\ }\bibfield  {title} {\bibinfo {title} {{Multi-messenger Observations of a Binary Neutron Star Merger}},\ }\href {https://doi.org/10.3847/2041-8213/aa91c9} {\bibfield  {journal} {\bibinfo  {journal} {Astrophys. J. Lett.}\ }\textbf {\bibinfo {volume} {848}},\ \bibinfo {eid} {L12} (\bibinfo {year} {2017})},\ \Eprint {https://arxiv.org/abs/1710.05833} {arXiv:1710.05833 [astro-ph.HE]} \BibitemShut {NoStop}%
\bibitem [{\citenamefont {{Goldstein}}\ \emph {et~al.}(2017)\citenamefont {{Goldstein}} \emph {et~al.}}]{Goldstein2017}%
  \BibitemOpen
  \bibfield  {author} {\bibinfo {author} {\bibfnamefont {A.}~\bibnamefont {{Goldstein}}} \emph {et~al.},\ }\bibfield  {title} {\bibinfo {title} {{An Ordinary Short Gamma-Ray Burst with Extraordinary Implications: Fermi-GBM Detection of GRB 170817A}},\ }\href {https://doi.org/10.3847/2041-8213/aa8f41} {\bibfield  {journal} {\bibinfo  {journal} {Astrophys. J., Lett.}\ }\textbf {\bibinfo {volume} {848}},\ \bibinfo {eid} {L14} (\bibinfo {year} {2017})},\ \Eprint {https://arxiv.org/abs/1710.05446} {arXiv:1710.05446 [astro-ph.HE]} \BibitemShut {NoStop}%
\bibitem [{\citenamefont {{Haggard}}\ \emph {et~al.}(2017)\citenamefont {{Haggard}}, \citenamefont {{Nynka}}, \citenamefont {{Ruan}}, \citenamefont {{Kalogera}}, \citenamefont {{Cenko}}, \citenamefont {{Evans}},\ and\ \citenamefont {{Kennea}}}]{Haggard2017}%
  \BibitemOpen
  \bibfield  {author} {\bibinfo {author} {\bibfnamefont {D.}~\bibnamefont {{Haggard}}}, \bibinfo {author} {\bibfnamefont {M.}~\bibnamefont {{Nynka}}}, \bibinfo {author} {\bibfnamefont {J.~J.}\ \bibnamefont {{Ruan}}}, \bibinfo {author} {\bibfnamefont {V.}~\bibnamefont {{Kalogera}}}, \bibinfo {author} {\bibfnamefont {S.~B.}\ \bibnamefont {{Cenko}}}, \bibinfo {author} {\bibfnamefont {P.}~\bibnamefont {{Evans}}},\ and\ \bibinfo {author} {\bibfnamefont {J.~A.}\ \bibnamefont {{Kennea}}},\ }\bibfield  {title} {\bibinfo {title} {{A Deep Chandra X-Ray Study of Neutron Star Coalescence GW170817}},\ }\href {https://doi.org/10.3847/2041-8213/aa8ede} {\bibfield  {journal} {\bibinfo  {journal} {\apjl}\ }\textbf {\bibinfo {volume} {848}},\ \bibinfo {eid} {L25} (\bibinfo {year} {2017})},\ \Eprint {https://arxiv.org/abs/1710.05852} {arXiv:1710.05852 [astro-ph.HE]} \BibitemShut {NoStop}%
\bibitem [{\citenamefont {{Abbott}}\ \emph {et~al.}(2017)\citenamefont {{Abbott}} \emph {et~al.}}]{GW_hubble}%
  \BibitemOpen
  \bibfield  {author} {\bibinfo {author} {\bibfnamefont {B.~P.}\ \bibnamefont {{Abbott}}} \emph {et~al.},\ }\bibfield  {title} {\bibinfo {title} {{A gravitational-wave standard siren measurement of the Hubble constant}},\ }\href {https://doi.org/10.1038/nature24471} {\bibfield  {journal} {\bibinfo  {journal} {Nature}\ }\textbf {\bibinfo {volume} {551}},\ \bibinfo {pages} {85} (\bibinfo {year} {2017})},\ \Eprint {https://arxiv.org/abs/1710.05835} {arXiv:1710.05835 [astro-ph.CO]} \BibitemShut {NoStop}%
\bibitem [{\citenamefont {Radice}\ \emph {et~al.}(2018)\citenamefont {Radice}, \citenamefont {Perego}, \citenamefont {Zappa},\ and\ \citenamefont {Bernuzzi}}]{Radice2018}%
  \BibitemOpen
  \bibfield  {author} {\bibinfo {author} {\bibfnamefont {D.}~\bibnamefont {Radice}}, \bibinfo {author} {\bibfnamefont {A.}~\bibnamefont {Perego}}, \bibinfo {author} {\bibfnamefont {F.}~\bibnamefont {Zappa}},\ and\ \bibinfo {author} {\bibfnamefont {S.}~\bibnamefont {Bernuzzi}},\ }\bibfield  {title} {\bibinfo {title} {{GW}170817: Joint constraint on the neutron star equation of state from multimessenger observations},\ }\href {https://doi.org/10.3847/2041-8213/aaa402} {\bibfield  {journal} {\bibinfo  {journal} {The Astrophysical Journal}\ }\textbf {\bibinfo {volume} {852}},\ \bibinfo {pages} {L29} (\bibinfo {year} {2018})}\BibitemShut {NoStop}%
\bibitem [{\citenamefont {Baiotti}(2019)}]{Baiotti2019}%
  \BibitemOpen
  \bibfield  {author} {\bibinfo {author} {\bibfnamefont {L.}~\bibnamefont {Baiotti}},\ }\bibfield  {title} {\bibinfo {title} {Gravitational waves from neutron star mergers and their relation to the nuclear equation of state},\ }\href {https://doi.org/https://doi.org/10.1016/j.ppnp.2019.103714} {\bibfield  {journal} {\bibinfo  {journal} {Progress in Particle and Nuclear Physics}\ }\textbf {\bibinfo {volume} {109}},\ \bibinfo {pages} {103714} (\bibinfo {year} {2019})}\BibitemShut {NoStop}%
\bibitem [{\citenamefont {Akutsu}\ \emph {et~al.}(2020)\citenamefont {Akutsu} \emph {et~al.}}]{KAGRA}%
  \BibitemOpen
  \bibfield  {author} {\bibinfo {author} {\bibfnamefont {T.}~\bibnamefont {Akutsu}} \emph {et~al.},\ }\bibfield  {title} {\bibinfo {title} {{Overview of KAGRA: Detector design and construction history}},\ }\href {https://doi.org/10.1093/ptep/ptaa125} {\bibfield  {journal} {\bibinfo  {journal} {Progress of Theoretical and Experimental Physics}\ }\textbf {\bibinfo {volume} {2021}},\ \bibinfo {pages} {05A101} (\bibinfo {year} {2020})},\ \Eprint {https://arxiv.org/abs/https://academic.oup.com/ptep/article-pdf/2021/5/05A101/37974994/ptaa125.pdf} {https://academic.oup.com/ptep/article-pdf/2021/5/05A101/37974994/ptaa125.pdf} \BibitemShut {NoStop}%
\bibitem [{\citenamefont {{Messick}}\ \emph {et~al.}(2017)\citenamefont {{Messick}} \emph {et~al.}}]{gstlal}%
  \BibitemOpen
  \bibfield  {author} {\bibinfo {author} {\bibfnamefont {C.}~\bibnamefont {{Messick}}} \emph {et~al.},\ }\bibfield  {title} {\bibinfo {title} {{Analysis framework for the prompt discovery of compact binary mergers in gravitational-wave data}},\ }\href {https://doi.org/10.1103/PhysRevD.95.042001} {\bibfield  {journal} {\bibinfo  {journal} {Physical Review D}\ }\textbf {\bibinfo {volume} {95}},\ \bibinfo {eid} {042001} (\bibinfo {year} {2017})},\ \Eprint {https://arxiv.org/abs/1604.04324} {arXiv:1604.04324 [astro-ph.IM]} \BibitemShut {NoStop}%
\bibitem [{\citenamefont {{Nitz}}\ \emph {et~al.}(2017)\citenamefont {{Nitz}}, \citenamefont {{Dent}}, \citenamefont {{Dal Canton}}, \citenamefont {{Fairhurst}},\ and\ \citenamefont {{Brown}}}]{pycbclive}%
  \BibitemOpen
  \bibfield  {author} {\bibinfo {author} {\bibfnamefont {A.~H.}\ \bibnamefont {{Nitz}}}, \bibinfo {author} {\bibfnamefont {T.}~\bibnamefont {{Dent}}}, \bibinfo {author} {\bibfnamefont {T.}~\bibnamefont {{Dal Canton}}}, \bibinfo {author} {\bibfnamefont {S.}~\bibnamefont {{Fairhurst}}},\ and\ \bibinfo {author} {\bibfnamefont {D.~A.}\ \bibnamefont {{Brown}}},\ }\bibfield  {title} {\bibinfo {title} {{Detecting Binary Compact-object Mergers with Gravitational Waves: Understanding and Improving the Sensitivity of the PyCBC Search}},\ }\href {https://doi.org/10.3847/1538-4357/aa8f50} {\bibfield  {journal} {\bibinfo  {journal} {Astrophys. J.}\ }\textbf {\bibinfo {volume} {849}},\ \bibinfo {eid} {118} (\bibinfo {year} {2017})},\ \Eprint {https://arxiv.org/abs/1705.01513} {arXiv:1705.01513 [gr-qc]} \BibitemShut {NoStop}%
\bibitem [{\citenamefont {{Adams}}\ \emph {et~al.}(2016)\citenamefont {{Adams}}, \citenamefont {{Buskulic}}, \citenamefont {{Germain}}, \citenamefont {{Guidi}}, \citenamefont {{Marion}}, \citenamefont {{Montani}}, \citenamefont {{Mours}}, \citenamefont {{Piergiovanni}},\ and\ \citenamefont {{Wang}}}]{mbta}%
  \BibitemOpen
  \bibfield  {author} {\bibinfo {author} {\bibfnamefont {T.}~\bibnamefont {{Adams}}}, \bibinfo {author} {\bibfnamefont {D.}~\bibnamefont {{Buskulic}}}, \bibinfo {author} {\bibfnamefont {V.}~\bibnamefont {{Germain}}}, \bibinfo {author} {\bibfnamefont {G.~M.}\ \bibnamefont {{Guidi}}}, \bibinfo {author} {\bibfnamefont {F.}~\bibnamefont {{Marion}}}, \bibinfo {author} {\bibfnamefont {M.}~\bibnamefont {{Montani}}}, \bibinfo {author} {\bibfnamefont {B.}~\bibnamefont {{Mours}}}, \bibinfo {author} {\bibfnamefont {F.}~\bibnamefont {{Piergiovanni}}},\ and\ \bibinfo {author} {\bibfnamefont {G.}~\bibnamefont {{Wang}}},\ }\bibfield  {title} {\bibinfo {title} {{Low-latency analysis pipeline for compact binary coalescences in the advanced gravitational wave detector era}},\ }\href {https://doi.org/10.1088/0264-9381/33/17/175012} {\bibfield  {journal} {\bibinfo  {journal} {Classical and Quantum Gravity}\ }\textbf {\bibinfo {volume} {33}},\ \bibinfo {eid} {175012} (\bibinfo {year} {2016})},\ \Eprint
  {https://arxiv.org/abs/1512.02864} {arXiv:1512.02864 [gr-qc]} \BibitemShut {NoStop}%
\bibitem [{\citenamefont {{Chu}}\ \emph {et~al.}(2022)\citenamefont {{Chu}}, \citenamefont {{Kovalam}}, \citenamefont {{Wen}}, \citenamefont {{Slaven-Blair}}, \citenamefont {{Bosveld}}, \citenamefont {{Chen}}, \citenamefont {{Clearwater}}, \citenamefont {{Codoreanu}}, \citenamefont {{Du}}, \citenamefont {{Guo}}, \citenamefont {{Guo}}, \citenamefont {{Kim}}, \citenamefont {{Li}}, \citenamefont {{Oloworaran}}, \citenamefont {{Panther}}, \citenamefont {{Powell}}, \citenamefont {{Sengupta}}, \citenamefont {{Wette}},\ and\ \citenamefont {{Zhu}}}]{Chu2022}%
  \BibitemOpen
  \bibfield  {author} {\bibinfo {author} {\bibfnamefont {Q.}~\bibnamefont {{Chu}}}, \bibinfo {author} {\bibfnamefont {M.}~\bibnamefont {{Kovalam}}}, \bibinfo {author} {\bibfnamefont {L.}~\bibnamefont {{Wen}}}, \bibinfo {author} {\bibfnamefont {T.}~\bibnamefont {{Slaven-Blair}}}, \bibinfo {author} {\bibfnamefont {J.}~\bibnamefont {{Bosveld}}}, \bibinfo {author} {\bibfnamefont {Y.}~\bibnamefont {{Chen}}}, \bibinfo {author} {\bibfnamefont {P.}~\bibnamefont {{Clearwater}}}, \bibinfo {author} {\bibfnamefont {A.}~\bibnamefont {{Codoreanu}}}, \bibinfo {author} {\bibfnamefont {Z.}~\bibnamefont {{Du}}}, \bibinfo {author} {\bibfnamefont {X.}~\bibnamefont {{Guo}}}, \bibinfo {author} {\bibfnamefont {X.}~\bibnamefont {{Guo}}}, \bibinfo {author} {\bibfnamefont {K.}~\bibnamefont {{Kim}}}, \bibinfo {author} {\bibfnamefont {T.~G.~F.}\ \bibnamefont {{Li}}}, \bibinfo {author} {\bibfnamefont {V.}~\bibnamefont {{Oloworaran}}}, \bibinfo {author} {\bibfnamefont {F.}~\bibnamefont {{Panther}}}, \bibinfo {author} {\bibfnamefont
  {J.}~\bibnamefont {{Powell}}}, \bibinfo {author} {\bibfnamefont {A.~S.}\ \bibnamefont {{Sengupta}}}, \bibinfo {author} {\bibfnamefont {K.}~\bibnamefont {{Wette}}},\ and\ \bibinfo {author} {\bibfnamefont {X.}~\bibnamefont {{Zhu}}},\ }\bibfield  {title} {\bibinfo {title} {{SPIIR online coherent pipeline to search for gravitational waves from compact binary coalescences}},\ }\href {https://doi.org/10.1103/PhysRevD.105.024023} {\bibfield  {journal} {\bibinfo  {journal} {\prd}\ }\textbf {\bibinfo {volume} {105}},\ \bibinfo {eid} {024023} (\bibinfo {year} {2022})},\ \Eprint {https://arxiv.org/abs/2109.14183} {arXiv:2109.14183 [gr-qc]} \BibitemShut {NoStop}%
\bibitem [{\citenamefont {{Klimenko}}\ \emph {et~al.}(2016)\citenamefont {{Klimenko}} \emph {et~al.}}]{cwb}%
  \BibitemOpen
  \bibfield  {author} {\bibinfo {author} {\bibfnamefont {S.}~\bibnamefont {{Klimenko}}} \emph {et~al.},\ }\bibfield  {title} {\bibinfo {title} {{Method for detection and reconstruction of gravitational wave transients with networks of advanced detectors}},\ }\href {https://doi.org/10.1103/PhysRevD.93.042004} {\bibfield  {journal} {\bibinfo  {journal} {Physical Review D}\ }\textbf {\bibinfo {volume} {93}},\ \bibinfo {eid} {042004} (\bibinfo {year} {2016})},\ \Eprint {https://arxiv.org/abs/1511.05999} {arXiv:1511.05999 [gr-qc]} \BibitemShut {NoStop}%
\bibitem [{\citenamefont {Allen}(2005)}]{Allen2005}%
  \BibitemOpen
  \bibfield  {author} {\bibinfo {author} {\bibfnamefont {B.}~\bibnamefont {Allen}},\ }\bibfield  {title} {\bibinfo {title} {${\ensuremath{\chi}}^{2}$ time-frequency discriminator for gravitational wave detection},\ }\href {https://doi.org/10.1103/PhysRevD.71.062001} {\bibfield  {journal} {\bibinfo  {journal} {Phys. Rev. D}\ }\textbf {\bibinfo {volume} {71}},\ \bibinfo {pages} {062001} (\bibinfo {year} {2005})}\BibitemShut {NoStop}%
\bibitem [{\citenamefont {{Dal Canton}}\ \emph {et~al.}(2021)\citenamefont {{Dal Canton}}, \citenamefont {{Nitz}}, \citenamefont {{Gadre}}, \citenamefont {{Cabourn Davies}}, \citenamefont {{Villa-Ortega}}, \citenamefont {{Dent}}, \citenamefont {{Harry}},\ and\ \citenamefont {{Xiao}}}]{DalCanton2021}%
  \BibitemOpen
  \bibfield  {author} {\bibinfo {author} {\bibfnamefont {T.}~\bibnamefont {{Dal Canton}}}, \bibinfo {author} {\bibfnamefont {A.~H.}\ \bibnamefont {{Nitz}}}, \bibinfo {author} {\bibfnamefont {B.}~\bibnamefont {{Gadre}}}, \bibinfo {author} {\bibfnamefont {G.~S.}\ \bibnamefont {{Cabourn Davies}}}, \bibinfo {author} {\bibfnamefont {V.}~\bibnamefont {{Villa-Ortega}}}, \bibinfo {author} {\bibfnamefont {T.}~\bibnamefont {{Dent}}}, \bibinfo {author} {\bibfnamefont {I.}~\bibnamefont {{Harry}}},\ and\ \bibinfo {author} {\bibfnamefont {L.}~\bibnamefont {{Xiao}}},\ }\bibfield  {title} {\bibinfo {title} {{Real-time Search for Compact Binary Mergers in Advanced LIGO and Virgo's Third Observing Run Using PyCBC Live}},\ }\href {https://doi.org/10.3847/1538-4357/ac2f9a} {\bibfield  {journal} {\bibinfo  {journal} {\apj}\ }\textbf {\bibinfo {volume} {923}},\ \bibinfo {eid} {254} (\bibinfo {year} {2021})},\ \Eprint {https://arxiv.org/abs/2008.07494} {arXiv:2008.07494 [astro-ph.HE]} \BibitemShut {NoStop}%
\bibitem [{\citenamefont {{Beveridge}}\ \emph {et~al.}(2025)\citenamefont {{Beveridge}}, \citenamefont {{McLeod}}, \citenamefont {{Wen}},\ and\ \citenamefont {{Wicenec}}}]{Beveridge2024}%
  \BibitemOpen
  \bibfield  {author} {\bibinfo {author} {\bibfnamefont {D.}~\bibnamefont {{Beveridge}}}, \bibinfo {author} {\bibfnamefont {A.}~\bibnamefont {{McLeod}}}, \bibinfo {author} {\bibfnamefont {L.}~\bibnamefont {{Wen}}},\ and\ \bibinfo {author} {\bibfnamefont {A.}~\bibnamefont {{Wicenec}}},\ }\bibfield  {title} {\bibinfo {title} {{Novel deep learning approach to detecting binary black hole mergers}},\ }\href {https://doi.org/10.1103/PhysRevD.111.024005} {\bibfield  {journal} {\bibinfo  {journal} {\prd}\ }\textbf {\bibinfo {volume} {111}},\ \bibinfo {eid} {024005} (\bibinfo {year} {2025})},\ \Eprint {https://arxiv.org/abs/2308.08429} {arXiv:2308.08429 [gr-qc]} \BibitemShut {NoStop}%
\bibitem [{\citenamefont {{Cuoco}}\ \emph {et~al.}(2021)\citenamefont {{Cuoco}} \emph {et~al.}}]{Cuoco2021}%
  \BibitemOpen
  \bibfield  {author} {\bibinfo {author} {\bibfnamefont {E.}~\bibnamefont {{Cuoco}}} \emph {et~al.},\ }\bibfield  {title} {\bibinfo {title} {{Enhancing gravitational-wave science with machine learning}},\ }\href {https://doi.org/10.1088/2632-2153/abb93a} {\bibfield  {journal} {\bibinfo  {journal} {Machine Learning: Science and Technology}\ }\textbf {\bibinfo {volume} {2}},\ \bibinfo {eid} {011002} (\bibinfo {year} {2021})},\ \Eprint {https://arxiv.org/abs/2005.03745} {arXiv:2005.03745 [astro-ph.HE]} \BibitemShut {NoStop}%
\bibitem [{\citenamefont {{Dax}}\ \emph {et~al.}(2021)\citenamefont {{Dax}}, \citenamefont {{Green}}, \citenamefont {{Gair}}, \citenamefont {{Macke}}, \citenamefont {{Buonanno}},\ and\ \citenamefont {{Sch{\"o}lkopf}}}]{Dax2021}%
  \BibitemOpen
  \bibfield  {author} {\bibinfo {author} {\bibfnamefont {M.}~\bibnamefont {{Dax}}}, \bibinfo {author} {\bibfnamefont {S.~R.}\ \bibnamefont {{Green}}}, \bibinfo {author} {\bibfnamefont {J.}~\bibnamefont {{Gair}}}, \bibinfo {author} {\bibfnamefont {J.~H.}\ \bibnamefont {{Macke}}}, \bibinfo {author} {\bibfnamefont {A.}~\bibnamefont {{Buonanno}}},\ and\ \bibinfo {author} {\bibfnamefont {B.}~\bibnamefont {{Sch{\"o}lkopf}}},\ }\bibfield  {title} {\bibinfo {title} {{Real-Time Gravitational Wave Science with Neural Posterior Estimation}},\ }\href {https://doi.org/10.1103/PhysRevLett.127.241103} {\bibfield  {journal} {\bibinfo  {journal} {\prl}\ }\textbf {\bibinfo {volume} {127}},\ \bibinfo {eid} {241103} (\bibinfo {year} {2021})},\ \Eprint {https://arxiv.org/abs/2106.12594} {arXiv:2106.12594 [gr-qc]} \BibitemShut {NoStop}%
\bibitem [{\citenamefont {{Gabbard}}\ \emph {et~al.}(2022)\citenamefont {{Gabbard}}, \citenamefont {{Messenger}}, \citenamefont {{Heng}}, \citenamefont {{Tonolini}},\ and\ \citenamefont {{Murray-Smith}}}]{Gabbard2022}%
  \BibitemOpen
  \bibfield  {author} {\bibinfo {author} {\bibfnamefont {H.}~\bibnamefont {{Gabbard}}}, \bibinfo {author} {\bibfnamefont {C.}~\bibnamefont {{Messenger}}}, \bibinfo {author} {\bibfnamefont {I.~S.}\ \bibnamefont {{Heng}}}, \bibinfo {author} {\bibfnamefont {F.}~\bibnamefont {{Tonolini}}},\ and\ \bibinfo {author} {\bibfnamefont {R.}~\bibnamefont {{Murray-Smith}}},\ }\bibfield  {title} {\bibinfo {title} {{Bayesian parameter estimation using conditional variational autoencoders for gravitational-wave astronomy}},\ }\href {https://doi.org/10.1038/s41567-021-01425-7} {\bibfield  {journal} {\bibinfo  {journal} {Nature Physics}\ }\textbf {\bibinfo {volume} {18}},\ \bibinfo {pages} {112} (\bibinfo {year} {2022})},\ \Eprint {https://arxiv.org/abs/1909.06296} {arXiv:1909.06296 [astro-ph.IM]} \BibitemShut {NoStop}%
\bibitem [{\citenamefont {{Chatterjee}}\ \emph {et~al.}(2023)\citenamefont {{Chatterjee}}, \citenamefont {{Kovalam}}, \citenamefont {{Wen}}, \citenamefont {{Beveridge}}, \citenamefont {{Diakogiannis}},\ and\ \citenamefont {{Vinsen}}}]{Chatterjee2023}%
  \BibitemOpen
  \bibfield  {author} {\bibinfo {author} {\bibfnamefont {C.}~\bibnamefont {{Chatterjee}}}, \bibinfo {author} {\bibfnamefont {M.}~\bibnamefont {{Kovalam}}}, \bibinfo {author} {\bibfnamefont {L.}~\bibnamefont {{Wen}}}, \bibinfo {author} {\bibfnamefont {D.}~\bibnamefont {{Beveridge}}}, \bibinfo {author} {\bibfnamefont {F.}~\bibnamefont {{Diakogiannis}}},\ and\ \bibinfo {author} {\bibfnamefont {K.}~\bibnamefont {{Vinsen}}},\ }\bibfield  {title} {\bibinfo {title} {{Rapid Localization of Gravitational Wave Sources from Compact Binary Coalescences Using Deep Learning}},\ }\href {https://doi.org/10.3847/1538-4357/ad08b7} {\bibfield  {journal} {\bibinfo  {journal} {\apj}\ }\textbf {\bibinfo {volume} {959}},\ \bibinfo {eid} {42} (\bibinfo {year} {2023})},\ \Eprint {https://arxiv.org/abs/2207.14522} {arXiv:2207.14522 [gr-qc]} \BibitemShut {NoStop}%
\bibitem [{\citenamefont {Bahaadini}\ \emph {et~al.}(2018)\citenamefont {Bahaadini}, \citenamefont {Noroozi}, \citenamefont {Rohani}, \citenamefont {Coughlin}, \citenamefont {Zevin}, \citenamefont {Smith}, \citenamefont {Kalogera},\ and\ \citenamefont {Katsaggelos}}]{Bahaadini2018}%
  \BibitemOpen
  \bibfield  {author} {\bibinfo {author} {\bibfnamefont {S.}~\bibnamefont {Bahaadini}}, \bibinfo {author} {\bibfnamefont {V.}~\bibnamefont {Noroozi}}, \bibinfo {author} {\bibfnamefont {N.}~\bibnamefont {Rohani}}, \bibinfo {author} {\bibfnamefont {S.}~\bibnamefont {Coughlin}}, \bibinfo {author} {\bibfnamefont {M.}~\bibnamefont {Zevin}}, \bibinfo {author} {\bibfnamefont {J.}~\bibnamefont {Smith}}, \bibinfo {author} {\bibfnamefont {V.}~\bibnamefont {Kalogera}},\ and\ \bibinfo {author} {\bibfnamefont {A.}~\bibnamefont {Katsaggelos}},\ }\bibfield  {title} {\bibinfo {title} {Machine learning for gravity spy: Glitch classification and dataset},\ }\href {https://doi.org/https://doi.org/10.1016/j.ins.2018.02.068} {\bibfield  {journal} {\bibinfo  {journal} {Information Sciences}\ }\textbf {\bibinfo {volume} {444}},\ \bibinfo {pages} {172} (\bibinfo {year} {2018})}\BibitemShut {NoStop}%
\bibitem [{\citenamefont {Essick}\ \emph {et~al.}(2020)\citenamefont {Essick}, \citenamefont {Godwin}, \citenamefont {Hanna}, \citenamefont {Blackburn},\ and\ \citenamefont {Katsavounidis}}]{Essick2020}%
  \BibitemOpen
  \bibfield  {author} {\bibinfo {author} {\bibfnamefont {R.}~\bibnamefont {Essick}}, \bibinfo {author} {\bibfnamefont {P.}~\bibnamefont {Godwin}}, \bibinfo {author} {\bibfnamefont {C.}~\bibnamefont {Hanna}}, \bibinfo {author} {\bibfnamefont {L.}~\bibnamefont {Blackburn}},\ and\ \bibinfo {author} {\bibfnamefont {E.}~\bibnamefont {Katsavounidis}},\ }\bibfield  {title} {\bibinfo {title} {{iDQ: Statistical inference of non-gaussian noise with auxiliary degrees of freedom in gravitational-wave detectors}},\ }\href {https://doi.org/10.1088/2632-2153/abab5f} {\bibfield  {journal} {\bibinfo  {journal} {Machine Learning: Science and Technology}\ }\textbf {\bibinfo {volume} {2}},\ \bibinfo {pages} {015004} (\bibinfo {year} {2020})}\BibitemShut {NoStop}%
\bibitem [{\citenamefont {Skliris}\ \emph {et~al.}(2024)\citenamefont {Skliris}, \citenamefont {Norman},\ and\ \citenamefont {Sutton}}]{Skliris2022}%
  \BibitemOpen
  \bibfield  {author} {\bibinfo {author} {\bibfnamefont {V.}~\bibnamefont {Skliris}}, \bibinfo {author} {\bibfnamefont {M.~R.~K.}\ \bibnamefont {Norman}},\ and\ \bibinfo {author} {\bibfnamefont {P.~J.}\ \bibnamefont {Sutton}},\ }\bibfield  {title} {\bibinfo {title} {Toward real-time detection of unmodeled gravitational wave transients using convolutional neural networks},\ }\href {https://doi.org/10.1103/PhysRevD.110.104034} {\bibfield  {journal} {\bibinfo  {journal} {Phys. Rev. D}\ }\textbf {\bibinfo {volume} {110}},\ \bibinfo {pages} {104034} (\bibinfo {year} {2024})}\BibitemShut {NoStop}%
\bibitem [{\citenamefont {{Sch{\"a}fer}}\ \emph {et~al.}(2023)\citenamefont {{Sch{\"a}fer}}, \citenamefont {{Zelenka}}, \citenamefont {{Nitz}}, \citenamefont {{Wang}}, \citenamefont {{Wu}}, \citenamefont {{Guo}}, \citenamefont {{Cao}}, \citenamefont {{Ren}}, \citenamefont {{Nousi}}, \citenamefont {{Stergioulas}}, \citenamefont {{Iosif}}, \citenamefont {{Koloniari}}, \citenamefont {{Tefas}}, \citenamefont {{Passalis}}, \citenamefont {{Salemi}}, \citenamefont {{Vedovato}}, \citenamefont {{Klimenko}}, \citenamefont {{Mishra}}, \citenamefont {{Br{\"u}gmann}}, \citenamefont {{Cuoco}}, \citenamefont {{Huerta}}, \citenamefont {{Messenger}},\ and\ \citenamefont {{Ohme}}}]{Schafer2023}%
  \BibitemOpen
  \bibfield  {author} {\bibinfo {author} {\bibfnamefont {M.~B.}\ \bibnamefont {{Sch{\"a}fer}}}, \bibinfo {author} {\bibfnamefont {O.}~\bibnamefont {{Zelenka}}}, \bibinfo {author} {\bibfnamefont {A.~H.}\ \bibnamefont {{Nitz}}}, \bibinfo {author} {\bibfnamefont {H.}~\bibnamefont {{Wang}}}, \bibinfo {author} {\bibfnamefont {S.}~\bibnamefont {{Wu}}}, \bibinfo {author} {\bibfnamefont {Z.-K.}\ \bibnamefont {{Guo}}}, \bibinfo {author} {\bibfnamefont {Z.}~\bibnamefont {{Cao}}}, \bibinfo {author} {\bibfnamefont {Z.}~\bibnamefont {{Ren}}}, \bibinfo {author} {\bibfnamefont {P.}~\bibnamefont {{Nousi}}}, \bibinfo {author} {\bibfnamefont {N.}~\bibnamefont {{Stergioulas}}}, \bibinfo {author} {\bibfnamefont {P.}~\bibnamefont {{Iosif}}}, \bibinfo {author} {\bibfnamefont {A.~E.}\ \bibnamefont {{Koloniari}}}, \bibinfo {author} {\bibfnamefont {A.}~\bibnamefont {{Tefas}}}, \bibinfo {author} {\bibfnamefont {N.}~\bibnamefont {{Passalis}}}, \bibinfo {author} {\bibfnamefont {F.}~\bibnamefont {{Salemi}}}, \bibinfo {author}
  {\bibfnamefont {G.}~\bibnamefont {{Vedovato}}}, \bibinfo {author} {\bibfnamefont {S.}~\bibnamefont {{Klimenko}}}, \bibinfo {author} {\bibfnamefont {T.}~\bibnamefont {{Mishra}}}, \bibinfo {author} {\bibfnamefont {B.}~\bibnamefont {{Br{\"u}gmann}}}, \bibinfo {author} {\bibfnamefont {E.}~\bibnamefont {{Cuoco}}}, \bibinfo {author} {\bibfnamefont {E.~A.}\ \bibnamefont {{Huerta}}}, \bibinfo {author} {\bibfnamefont {C.}~\bibnamefont {{Messenger}}},\ and\ \bibinfo {author} {\bibfnamefont {F.}~\bibnamefont {{Ohme}}},\ }\bibfield  {title} {\bibinfo {title} {{First machine learning gravitational-wave search mock data challenge}},\ }\href {https://doi.org/10.1103/PhysRevD.107.023021} {\bibfield  {journal} {\bibinfo  {journal} {\prd}\ }\textbf {\bibinfo {volume} {107}},\ \bibinfo {eid} {023021} (\bibinfo {year} {2023})},\ \Eprint {https://arxiv.org/abs/2209.11146} {arXiv:2209.11146 [astro-ph.IM]} \BibitemShut {NoStop}%
\bibitem [{\citenamefont {{Nousi}}\ \emph {et~al.}(2023)\citenamefont {{Nousi}}, \citenamefont {{Koloniari}}, \citenamefont {{Passalis}}, \citenamefont {{Iosif}}, \citenamefont {{Stergioulas}},\ and\ \citenamefont {{Tefas}}}]{Nousi2023}%
  \BibitemOpen
  \bibfield  {author} {\bibinfo {author} {\bibfnamefont {P.}~\bibnamefont {{Nousi}}}, \bibinfo {author} {\bibfnamefont {A.~E.}\ \bibnamefont {{Koloniari}}}, \bibinfo {author} {\bibfnamefont {N.}~\bibnamefont {{Passalis}}}, \bibinfo {author} {\bibfnamefont {P.}~\bibnamefont {{Iosif}}}, \bibinfo {author} {\bibfnamefont {N.}~\bibnamefont {{Stergioulas}}},\ and\ \bibinfo {author} {\bibfnamefont {A.}~\bibnamefont {{Tefas}}},\ }\bibfield  {title} {\bibinfo {title} {{Deep residual networks for gravitational wave detection}},\ }\href {https://doi.org/10.1103/PhysRevD.108.024022} {\bibfield  {journal} {\bibinfo  {journal} {\prd}\ }\textbf {\bibinfo {volume} {108}},\ \bibinfo {eid} {024022} (\bibinfo {year} {2023})},\ \Eprint {https://arxiv.org/abs/2211.01520} {arXiv:2211.01520 [gr-qc]} \BibitemShut {NoStop}%
\bibitem [{\citenamefont {{Marx}}\ \emph {et~al.}(2024)\citenamefont {{Marx}}, \citenamefont {{Benoit}}, \citenamefont {{Gunny}}, \citenamefont {{Omer}}, \citenamefont {{Chatterjee}}, \citenamefont {{Venterea}}, \citenamefont {{Wills}}, \citenamefont {{Saleem}}, \citenamefont {{Moreno}}, \citenamefont {{Raikman}}, \citenamefont {{Govorkova}}, \citenamefont {{Rankin}}, \citenamefont {{Coughlin}}, \citenamefont {{Harris}},\ and\ \citenamefont {{Katsavounidis}}}]{Marx2024}%
  \BibitemOpen
  \bibfield  {author} {\bibinfo {author} {\bibfnamefont {E.}~\bibnamefont {{Marx}}}, \bibinfo {author} {\bibfnamefont {W.}~\bibnamefont {{Benoit}}}, \bibinfo {author} {\bibfnamefont {A.}~\bibnamefont {{Gunny}}}, \bibinfo {author} {\bibfnamefont {R.}~\bibnamefont {{Omer}}}, \bibinfo {author} {\bibfnamefont {D.}~\bibnamefont {{Chatterjee}}}, \bibinfo {author} {\bibfnamefont {R.~C.}\ \bibnamefont {{Venterea}}}, \bibinfo {author} {\bibfnamefont {L.}~\bibnamefont {{Wills}}}, \bibinfo {author} {\bibfnamefont {M.}~\bibnamefont {{Saleem}}}, \bibinfo {author} {\bibfnamefont {E.}~\bibnamefont {{Moreno}}}, \bibinfo {author} {\bibfnamefont {R.}~\bibnamefont {{Raikman}}}, \bibinfo {author} {\bibfnamefont {E.}~\bibnamefont {{Govorkova}}}, \bibinfo {author} {\bibfnamefont {D.}~\bibnamefont {{Rankin}}}, \bibinfo {author} {\bibfnamefont {M.~W.}\ \bibnamefont {{Coughlin}}}, \bibinfo {author} {\bibfnamefont {P.}~\bibnamefont {{Harris}}},\ and\ \bibinfo {author} {\bibfnamefont {E.}~\bibnamefont {{Katsavounidis}}},\ }\bibfield
  {title} {\bibinfo {title} {{A machine-learning pipeline for real-time detection of gravitational waves from compact binary coalescences}},\ }\bibfield  {journal} {\bibinfo  {journal} {arXiv e-prints}\ }\href {https://doi.org/10.48550/arXiv.2403.18661} {10.48550/arXiv.2403.18661} (\bibinfo {year} {2024}),\ \Eprint {https://arxiv.org/abs/2403.18661} {arXiv:2403.18661 [gr-qc]} \BibitemShut {NoStop}%
\bibitem [{\citenamefont {{Krastev}}(2020)}]{Krastev2020}%
  \BibitemOpen
  \bibfield  {author} {\bibinfo {author} {\bibfnamefont {P.~G.}\ \bibnamefont {{Krastev}}},\ }\bibfield  {title} {\bibinfo {title} {Real-time detection of gravitational waves from binary neutron stars using artificial neural networks},\ }\href {https://doi.org/10.1016/j.physletb.2020.135330} {\bibfield  {journal} {\bibinfo  {journal} {Physics Letters B}\ }\textbf {\bibinfo {volume} {803}},\ \bibinfo {pages} {135330} (\bibinfo {year} {2020})}\BibitemShut {NoStop}%
\bibitem [{\citenamefont {{Krastev}}\ \emph {et~al.}(2021)\citenamefont {{Krastev}}, \citenamefont {{Gill}}, \citenamefont {{Villar}},\ and\ \citenamefont {{Berger}}}]{Krastev2021}%
  \BibitemOpen
  \bibfield  {author} {\bibinfo {author} {\bibfnamefont {P.~G.}\ \bibnamefont {{Krastev}}}, \bibinfo {author} {\bibfnamefont {K.}~\bibnamefont {{Gill}}}, \bibinfo {author} {\bibfnamefont {V.~A.}\ \bibnamefont {{Villar}}},\ and\ \bibinfo {author} {\bibfnamefont {E.}~\bibnamefont {{Berger}}},\ }\bibfield  {title} {\bibinfo {title} {{Detection and parameter estimation of gravitational waves from binary neutron-star mergers in real LIGO data using deep learning}},\ }\href {https://doi.org/10.1016/j.physletb.2021.136161} {\bibfield  {journal} {\bibinfo  {journal} {Physics Letters B}\ }\textbf {\bibinfo {volume} {815}},\ \bibinfo {eid} {136161} (\bibinfo {year} {2021})},\ \Eprint {https://arxiv.org/abs/2012.13101} {arXiv:2012.13101 [astro-ph.IM]} \BibitemShut {NoStop}%
\bibitem [{\citenamefont {{Sch{\"a}fer}}\ \emph {et~al.}(2020)\citenamefont {{Sch{\"a}fer}}, \citenamefont {{Ohme}},\ and\ \citenamefont {{Nitz}}}]{Schafer2020}%
  \BibitemOpen
  \bibfield  {author} {\bibinfo {author} {\bibfnamefont {M.~B.}\ \bibnamefont {{Sch{\"a}fer}}}, \bibinfo {author} {\bibfnamefont {F.}~\bibnamefont {{Ohme}}},\ and\ \bibinfo {author} {\bibfnamefont {A.~H.}\ \bibnamefont {{Nitz}}},\ }\bibfield  {title} {\bibinfo {title} {{Detection of gravitational-wave signals from binary neutron star mergers using machine learning}},\ }\href {https://doi.org/10.1103/PhysRevD.102.063015} {\bibfield  {journal} {\bibinfo  {journal} {Physical Review D}\ }\textbf {\bibinfo {volume} {102}},\ \bibinfo {eid} {063015} (\bibinfo {year} {2020})},\ \Eprint {https://arxiv.org/abs/2006.01509} {arXiv:2006.01509 [astro-ph.HE]} \BibitemShut {NoStop}%
\bibitem [{\citenamefont {{Baltus}}\ \emph {et~al.}(2021)\citenamefont {{Baltus}}, \citenamefont {{Janquart}}, \citenamefont {{Lopez}}, \citenamefont {{Reza}}, \citenamefont {{Caudill}},\ and\ \citenamefont {{Cudell}}}]{Baltus2021}%
  \BibitemOpen
  \bibfield  {author} {\bibinfo {author} {\bibfnamefont {G.}~\bibnamefont {{Baltus}}}, \bibinfo {author} {\bibfnamefont {J.}~\bibnamefont {{Janquart}}}, \bibinfo {author} {\bibfnamefont {M.}~\bibnamefont {{Lopez}}}, \bibinfo {author} {\bibfnamefont {A.}~\bibnamefont {{Reza}}}, \bibinfo {author} {\bibfnamefont {S.}~\bibnamefont {{Caudill}}},\ and\ \bibinfo {author} {\bibfnamefont {J.-R.}\ \bibnamefont {{Cudell}}},\ }\bibfield  {title} {\bibinfo {title} {{Convolutional neural networks for the detection of the early inspiral of a gravitational-wave signal}},\ }\href {https://doi.org/10.1103/PhysRevD.103.102003} {\bibfield  {journal} {\bibinfo  {journal} {\prd}\ }\textbf {\bibinfo {volume} {103}},\ \bibinfo {eid} {102003} (\bibinfo {year} {2021})},\ \Eprint {https://arxiv.org/abs/2104.00594} {arXiv:2104.00594 [gr-qc]} \BibitemShut {NoStop}%
\bibitem [{\citenamefont {{Qiu}}\ \emph {et~al.}(2023)\citenamefont {{Qiu}}, \citenamefont {{Krastev}}, \citenamefont {{Gill}},\ and\ \citenamefont {{Berger}}}]{Qiu2023}%
  \BibitemOpen
  \bibfield  {author} {\bibinfo {author} {\bibfnamefont {R.}~\bibnamefont {{Qiu}}}, \bibinfo {author} {\bibfnamefont {P.~G.}\ \bibnamefont {{Krastev}}}, \bibinfo {author} {\bibfnamefont {K.}~\bibnamefont {{Gill}}},\ and\ \bibinfo {author} {\bibfnamefont {E.}~\bibnamefont {{Berger}}},\ }\bibfield  {title} {\bibinfo {title} {{Deep learning detection and classification of gravitational waves from neutron star-black hole mergers}},\ }\href {https://doi.org/10.1016/j.physletb.2023.137850} {\bibfield  {journal} {\bibinfo  {journal} {Physics Letters B}\ }\textbf {\bibinfo {volume} {840}},\ \bibinfo {eid} {137850} (\bibinfo {year} {2023})},\ \Eprint {https://arxiv.org/abs/2210.15888} {arXiv:2210.15888 [astro-ph.IM]} \BibitemShut {NoStop}%
\bibitem [{\citenamefont {{Wei}}\ and\ \citenamefont {{Huerta}}(2021)}]{Wei2021}%
  \BibitemOpen
  \bibfield  {author} {\bibinfo {author} {\bibfnamefont {W.}~\bibnamefont {{Wei}}}\ and\ \bibinfo {author} {\bibfnamefont {E.~A.}\ \bibnamefont {{Huerta}}},\ }\bibfield  {title} {\bibinfo {title} {{Deep learning for gravitational wave forecasting of neutron star mergers}},\ }\href {https://doi.org/10.1016/j.physletb.2021.136185} {\bibfield  {journal} {\bibinfo  {journal} {Physics Letters B}\ }\textbf {\bibinfo {volume} {816}},\ \bibinfo {eid} {136185} (\bibinfo {year} {2021})},\ \Eprint {https://arxiv.org/abs/2010.09751} {arXiv:2010.09751 [gr-qc]} \BibitemShut {NoStop}%
\bibitem [{\citenamefont {{Aveiro}}\ \emph {et~al.}(2022)\citenamefont {{Aveiro}}, \citenamefont {{Freitas}}, \citenamefont {{Ferreira}}, \citenamefont {{Onofre}}, \citenamefont {{Provid{\^e}ncia}}, \citenamefont {{Gon{\c{c}}alves}},\ and\ \citenamefont {{Font}}}]{Aveiro2022}%
  \BibitemOpen
  \bibfield  {author} {\bibinfo {author} {\bibfnamefont {J.}~\bibnamefont {{Aveiro}}}, \bibinfo {author} {\bibfnamefont {F.~F.}\ \bibnamefont {{Freitas}}}, \bibinfo {author} {\bibfnamefont {M.}~\bibnamefont {{Ferreira}}}, \bibinfo {author} {\bibfnamefont {A.}~\bibnamefont {{Onofre}}}, \bibinfo {author} {\bibfnamefont {C.}~\bibnamefont {{Provid{\^e}ncia}}}, \bibinfo {author} {\bibfnamefont {G.}~\bibnamefont {{Gon{\c{c}}alves}}},\ and\ \bibinfo {author} {\bibfnamefont {J.~A.}\ \bibnamefont {{Font}}},\ }\bibfield  {title} {\bibinfo {title} {{Identification of binary neutron star mergers in gravitational-wave data using object-detection machine learning models}},\ }\href {https://doi.org/10.1103/PhysRevD.106.084059} {\bibfield  {journal} {\bibinfo  {journal} {\prd}\ }\textbf {\bibinfo {volume} {106}},\ \bibinfo {eid} {084059} (\bibinfo {year} {2022})},\ \Eprint {https://arxiv.org/abs/2207.00591} {arXiv:2207.00591 [astro-ph.IM]} \BibitemShut {NoStop}%
\bibitem [{\citenamefont {{Allen}}\ \emph {et~al.}(2012)\citenamefont {{Allen}}, \citenamefont {{Anderson}}, \citenamefont {{Brady}}, \citenamefont {{Brown}},\ and\ \citenamefont {{Creighton}}}]{Allen2012}%
  \BibitemOpen
  \bibfield  {author} {\bibinfo {author} {\bibfnamefont {B.}~\bibnamefont {{Allen}}}, \bibinfo {author} {\bibfnamefont {W.~G.}\ \bibnamefont {{Anderson}}}, \bibinfo {author} {\bibfnamefont {P.~R.}\ \bibnamefont {{Brady}}}, \bibinfo {author} {\bibfnamefont {D.~A.}\ \bibnamefont {{Brown}}},\ and\ \bibinfo {author} {\bibfnamefont {J.~D.~E.}\ \bibnamefont {{Creighton}}},\ }\bibfield  {title} {\bibinfo {title} {{FINDCHIRP: An algorithm for detection of gravitational waves from inspiraling compact binaries}},\ }\href {https://doi.org/10.1103/PhysRevD.85.122006} {\bibfield  {journal} {\bibinfo  {journal} {\prd}\ }\textbf {\bibinfo {volume} {85}},\ \bibinfo {eid} {122006} (\bibinfo {year} {2012})},\ \Eprint {https://arxiv.org/abs/gr-qc/0509116} {arXiv:gr-qc/0509116 [gr-qc]} \BibitemShut {NoStop}%
\bibitem [{\citenamefont {Nitz}\ \emph {et~al.}(2018)\citenamefont {Nitz}, \citenamefont {Dal~Canton}, \citenamefont {Davis},\ and\ \citenamefont {Reyes}}]{Nitz2018}%
  \BibitemOpen
  \bibfield  {author} {\bibinfo {author} {\bibfnamefont {A.~H.}\ \bibnamefont {Nitz}}, \bibinfo {author} {\bibfnamefont {T.}~\bibnamefont {Dal~Canton}}, \bibinfo {author} {\bibfnamefont {D.}~\bibnamefont {Davis}},\ and\ \bibinfo {author} {\bibfnamefont {S.}~\bibnamefont {Reyes}},\ }\bibfield  {title} {\bibinfo {title} {{Rapid detection of gravitational waves from compact binary mergers with PyCBC Live}},\ }\href {https://doi.org/10.1103/PhysRevD.98.024050} {\bibfield  {journal} {\bibinfo  {journal} {Phys. Rev. D}\ }\textbf {\bibinfo {volume} {98}},\ \bibinfo {pages} {024050} (\bibinfo {year} {2018})}\BibitemShut {NoStop}%
\bibitem [{\citenamefont {Harris}\ \emph {et~al.}(2020)\citenamefont {Harris}, \citenamefont {Millman}, \citenamefont {van~der Walt}, \citenamefont {Gommers}, \citenamefont {Virtanen}, \citenamefont {Cournapeau}, \citenamefont {Wieser}, \citenamefont {Taylor}, \citenamefont {Berg}, \citenamefont {Smith}, \citenamefont {Kern}, \citenamefont {Picus}, \citenamefont {Hoyer}, \citenamefont {van Kerkwijk}, \citenamefont {Brett}, \citenamefont {Haldane}, \citenamefont {del R{\'{i}}o}, \citenamefont {Wiebe}, \citenamefont {Peterson}, \citenamefont {G{\'{e}}rard-Marchant}, \citenamefont {Sheppard}, \citenamefont {Reddy}, \citenamefont {Weckesser}, \citenamefont {Abbasi}, \citenamefont {Gohlke},\ and\ \citenamefont {Oliphant}}]{Harris2020}%
  \BibitemOpen
  \bibfield  {author} {\bibinfo {author} {\bibfnamefont {C.~R.}\ \bibnamefont {Harris}}, \bibinfo {author} {\bibfnamefont {K.~J.}\ \bibnamefont {Millman}}, \bibinfo {author} {\bibfnamefont {S.~J.}\ \bibnamefont {van~der Walt}}, \bibinfo {author} {\bibfnamefont {R.}~\bibnamefont {Gommers}}, \bibinfo {author} {\bibfnamefont {P.}~\bibnamefont {Virtanen}}, \bibinfo {author} {\bibfnamefont {D.}~\bibnamefont {Cournapeau}}, \bibinfo {author} {\bibfnamefont {E.}~\bibnamefont {Wieser}}, \bibinfo {author} {\bibfnamefont {J.}~\bibnamefont {Taylor}}, \bibinfo {author} {\bibfnamefont {S.}~\bibnamefont {Berg}}, \bibinfo {author} {\bibfnamefont {N.~J.}\ \bibnamefont {Smith}}, \bibinfo {author} {\bibfnamefont {R.}~\bibnamefont {Kern}}, \bibinfo {author} {\bibfnamefont {M.}~\bibnamefont {Picus}}, \bibinfo {author} {\bibfnamefont {S.}~\bibnamefont {Hoyer}}, \bibinfo {author} {\bibfnamefont {M.~H.}\ \bibnamefont {van Kerkwijk}}, \bibinfo {author} {\bibfnamefont {M.}~\bibnamefont {Brett}}, \bibinfo {author} {\bibfnamefont
  {A.}~\bibnamefont {Haldane}}, \bibinfo {author} {\bibfnamefont {J.~F.}\ \bibnamefont {del R{\'{i}}o}}, \bibinfo {author} {\bibfnamefont {M.}~\bibnamefont {Wiebe}}, \bibinfo {author} {\bibfnamefont {P.}~\bibnamefont {Peterson}}, \bibinfo {author} {\bibfnamefont {P.}~\bibnamefont {G{\'{e}}rard-Marchant}}, \bibinfo {author} {\bibfnamefont {K.}~\bibnamefont {Sheppard}}, \bibinfo {author} {\bibfnamefont {T.}~\bibnamefont {Reddy}}, \bibinfo {author} {\bibfnamefont {W.}~\bibnamefont {Weckesser}}, \bibinfo {author} {\bibfnamefont {H.}~\bibnamefont {Abbasi}}, \bibinfo {author} {\bibfnamefont {C.}~\bibnamefont {Gohlke}},\ and\ \bibinfo {author} {\bibfnamefont {T.~E.}\ \bibnamefont {Oliphant}},\ }\bibfield  {title} {\bibinfo {title} {Array programming with {NumPy}},\ }\href {https://doi.org/10.1038/s41586-020-2649-2} {\bibfield  {journal} {\bibinfo  {journal} {Nature}\ }\textbf {\bibinfo {volume} {585}},\ \bibinfo {pages} {357} (\bibinfo {year} {2020})}\BibitemShut {NoStop}%
\bibitem [{\citenamefont {Nitz}\ \emph {et~al.}(2023)\citenamefont {Nitz}, \citenamefont {Harry}, \citenamefont {Brown}, \citenamefont {Biwer}, \citenamefont {Willis}, \citenamefont {Canton}, \citenamefont {Capano}, \citenamefont {Dent}, \citenamefont {Pekowsky}, \citenamefont {Davies}, \citenamefont {De}, \citenamefont {Cabero}, \citenamefont {Wu}, \citenamefont {Williamson}, \citenamefont {Macleod}, \citenamefont {Machenschalk}, \citenamefont {Pannarale}, \citenamefont {Kumar}, \citenamefont {Reyes}, \citenamefont {dfinstad}, \citenamefont {Kumar}, \citenamefont {Tápai}, \citenamefont {Singer}, \citenamefont {Kumar}, \citenamefont {Gadre}, \citenamefont {maxtrevor}, \citenamefont {veronica villa}, \citenamefont {Khan}, \citenamefont {Fairhurst},\ and\ \citenamefont {Chandra}}]{pycbc_software}%
  \BibitemOpen
  \bibfield  {author} {\bibinfo {author} {\bibfnamefont {A.}~\bibnamefont {Nitz}}, \bibinfo {author} {\bibfnamefont {I.}~\bibnamefont {Harry}}, \bibinfo {author} {\bibfnamefont {D.}~\bibnamefont {Brown}}, \bibinfo {author} {\bibfnamefont {C.~M.}\ \bibnamefont {Biwer}}, \bibinfo {author} {\bibfnamefont {J.}~\bibnamefont {Willis}}, \bibinfo {author} {\bibfnamefont {T.~D.}\ \bibnamefont {Canton}}, \bibinfo {author} {\bibfnamefont {C.}~\bibnamefont {Capano}}, \bibinfo {author} {\bibfnamefont {T.}~\bibnamefont {Dent}}, \bibinfo {author} {\bibfnamefont {L.}~\bibnamefont {Pekowsky}}, \bibinfo {author} {\bibfnamefont {G.~S.~C.}\ \bibnamefont {Davies}}, \bibinfo {author} {\bibfnamefont {S.}~\bibnamefont {De}}, \bibinfo {author} {\bibfnamefont {M.}~\bibnamefont {Cabero}}, \bibinfo {author} {\bibfnamefont {S.}~\bibnamefont {Wu}}, \bibinfo {author} {\bibfnamefont {A.~R.}\ \bibnamefont {Williamson}}, \bibinfo {author} {\bibfnamefont {D.}~\bibnamefont {Macleod}}, \bibinfo {author} {\bibfnamefont {B.}~\bibnamefont
  {Machenschalk}}, \bibinfo {author} {\bibfnamefont {F.}~\bibnamefont {Pannarale}}, \bibinfo {author} {\bibfnamefont {P.}~\bibnamefont {Kumar}}, \bibinfo {author} {\bibfnamefont {S.}~\bibnamefont {Reyes}}, \bibinfo {author} {\bibnamefont {dfinstad}}, \bibinfo {author} {\bibfnamefont {S.}~\bibnamefont {Kumar}}, \bibinfo {author} {\bibfnamefont {M.}~\bibnamefont {Tápai}}, \bibinfo {author} {\bibfnamefont {L.}~\bibnamefont {Singer}}, \bibinfo {author} {\bibfnamefont {P.}~\bibnamefont {Kumar}}, \bibinfo {author} {\bibfnamefont {B.~U.~V.}\ \bibnamefont {Gadre}}, \bibinfo {author} {\bibnamefont {maxtrevor}}, \bibinfo {author} {\bibnamefont {veronica villa}}, \bibinfo {author} {\bibfnamefont {S.}~\bibnamefont {Khan}}, \bibinfo {author} {\bibfnamefont {S.}~\bibnamefont {Fairhurst}},\ and\ \bibinfo {author} {\bibfnamefont {K.}~\bibnamefont {Chandra}},\ }\href {https://doi.org/10.5281/zenodo.10137381} {\bibinfo {title} {gwastro/pycbc: v2.3.2 release of pycbc}} (\bibinfo {year} {2023})\BibitemShut {NoStop}%
\bibitem [{\citenamefont {{Brown}}\ \emph {et~al.}(2012)\citenamefont {{Brown}}, \citenamefont {{Harry}}, \citenamefont {{Lundgren}},\ and\ \citenamefont {{Nitz}}}]{Brown2012}%
  \BibitemOpen
  \bibfield  {author} {\bibinfo {author} {\bibfnamefont {D.~A.}\ \bibnamefont {{Brown}}}, \bibinfo {author} {\bibfnamefont {I.}~\bibnamefont {{Harry}}}, \bibinfo {author} {\bibfnamefont {A.}~\bibnamefont {{Lundgren}}},\ and\ \bibinfo {author} {\bibfnamefont {A.~H.}\ \bibnamefont {{Nitz}}},\ }\bibfield  {title} {\bibinfo {title} {{Detecting binary neutron star systems with spin in advanced gravitational-wave detectors}},\ }\href {https://doi.org/10.1103/PhysRevD.86.084017} {\bibfield  {journal} {\bibinfo  {journal} {\prd}\ }\textbf {\bibinfo {volume} {86}},\ \bibinfo {eid} {084017} (\bibinfo {year} {2012})},\ \Eprint {https://arxiv.org/abs/1207.6406} {arXiv:1207.6406 [gr-qc]} \BibitemShut {NoStop}%
\bibitem [{\citenamefont {{Buonanno}}\ \emph {et~al.}(2009)\citenamefont {{Buonanno}}, \citenamefont {{Iyer}}, \citenamefont {{Ochsner}}, \citenamefont {{Pan}},\ and\ \citenamefont {{Sathyaprakash}}}]{Buonanno2009}%
  \BibitemOpen
  \bibfield  {author} {\bibinfo {author} {\bibfnamefont {A.}~\bibnamefont {{Buonanno}}}, \bibinfo {author} {\bibfnamefont {B.~R.}\ \bibnamefont {{Iyer}}}, \bibinfo {author} {\bibfnamefont {E.}~\bibnamefont {{Ochsner}}}, \bibinfo {author} {\bibfnamefont {Y.}~\bibnamefont {{Pan}}},\ and\ \bibinfo {author} {\bibfnamefont {B.~S.}\ \bibnamefont {{Sathyaprakash}}},\ }\bibfield  {title} {\bibinfo {title} {{Comparison of post-Newtonian templates for compact binary inspiral signals in gravitational-wave detectors}},\ }\href {https://doi.org/10.1103/PhysRevD.80.084043} {\bibfield  {journal} {\bibinfo  {journal} {\prd}\ }\textbf {\bibinfo {volume} {80}},\ \bibinfo {eid} {084043} (\bibinfo {year} {2009})},\ \Eprint {https://arxiv.org/abs/0907.0700} {arXiv:0907.0700 [gr-qc]} \BibitemShut {NoStop}%
\bibitem [{\citenamefont {Harry}\ \emph {et~al.}(2009)\citenamefont {Harry}, \citenamefont {Allen},\ and\ \citenamefont {Sathyaprakash}}]{Harry2009}%
  \BibitemOpen
  \bibfield  {author} {\bibinfo {author} {\bibfnamefont {I.~W.}\ \bibnamefont {Harry}}, \bibinfo {author} {\bibfnamefont {B.}~\bibnamefont {Allen}},\ and\ \bibinfo {author} {\bibfnamefont {B.~S.}\ \bibnamefont {Sathyaprakash}},\ }\bibfield  {title} {\bibinfo {title} {Stochastic template placement algorithm for gravitational wave data analysis},\ }\href {https://doi.org/10.1103/PhysRevD.80.104014} {\bibfield  {journal} {\bibinfo  {journal} {Phys. Rev. D}\ }\textbf {\bibinfo {volume} {80}},\ \bibinfo {pages} {104014} (\bibinfo {year} {2009})}\BibitemShut {NoStop}%
\bibitem [{\citenamefont {Zhu}\ \emph {et~al.}(2018)\citenamefont {Zhu}, \citenamefont {Thrane}, \citenamefont {Os\l{}owski}, \citenamefont {Levin},\ and\ \citenamefont {Lasky}}]{Zhu2018}%
  \BibitemOpen
  \bibfield  {author} {\bibinfo {author} {\bibfnamefont {X.}~\bibnamefont {Zhu}}, \bibinfo {author} {\bibfnamefont {E.}~\bibnamefont {Thrane}}, \bibinfo {author} {\bibfnamefont {S.}~\bibnamefont {Os\l{}owski}}, \bibinfo {author} {\bibfnamefont {Y.}~\bibnamefont {Levin}},\ and\ \bibinfo {author} {\bibfnamefont {P.~D.}\ \bibnamefont {Lasky}},\ }\bibfield  {title} {\bibinfo {title} {Inferring the population properties of binary neutron stars with gravitational-wave measurements of spin},\ }\href {https://doi.org/10.1103/PhysRevD.98.043002} {\bibfield  {journal} {\bibinfo  {journal} {Phys. Rev. D}\ }\textbf {\bibinfo {volume} {98}},\ \bibinfo {pages} {043002} (\bibinfo {year} {2018})}\BibitemShut {NoStop}%
\bibitem [{\citenamefont {Abbott}\ \emph {et~al.}(2023{\natexlab{b}})\citenamefont {Abbott} \emph {et~al.}}]{gwosc}%
  \BibitemOpen
  \bibfield  {author} {\bibinfo {author} {\bibfnamefont {R.}~\bibnamefont {Abbott}} \emph {et~al.} (\bibinfo {collaboration} {LIGO Scientific Collaboration and Virgo Collaboration and KAGRA Collaboration}),\ }\bibfield  {title} {\bibinfo {title} {{Open Data from the Third Observing Run of LIGO, Virgo, KAGRA, and GEO}},\ }\href {https://doi.org/10.3847/1538-4365/acdc9f} {\bibfield  {journal} {\bibinfo  {journal} {Astrophys. J. Suppl.}\ }\textbf {\bibinfo {volume} {267}},\ \bibinfo {pages} {29} (\bibinfo {year} {2023}{\natexlab{b}})},\ \Eprint {https://arxiv.org/abs/2302.03676} {arXiv:2302.03676 [gr-qc]} \BibitemShut {NoStop}%
\bibitem [{\citenamefont {{Robinet}}\ \emph {et~al.}(2020)\citenamefont {{Robinet}}, \citenamefont {{Arnaud}}, \citenamefont {{Leroy}}, \citenamefont {{Lundgren}}, \citenamefont {{Macleod}},\ and\ \citenamefont {{McIver}}}]{Robinet2020}%
  \BibitemOpen
  \bibfield  {author} {\bibinfo {author} {\bibfnamefont {F.}~\bibnamefont {{Robinet}}}, \bibinfo {author} {\bibfnamefont {N.}~\bibnamefont {{Arnaud}}}, \bibinfo {author} {\bibfnamefont {N.}~\bibnamefont {{Leroy}}}, \bibinfo {author} {\bibfnamefont {A.}~\bibnamefont {{Lundgren}}}, \bibinfo {author} {\bibfnamefont {D.}~\bibnamefont {{Macleod}}},\ and\ \bibinfo {author} {\bibfnamefont {J.}~\bibnamefont {{McIver}}},\ }\bibfield  {title} {\bibinfo {title} {{Omicron: A tool to characterize transient noise in gravitational-wave detectors}},\ }\href {https://doi.org/10.1016/j.softx.2020.100620} {\bibfield  {journal} {\bibinfo  {journal} {SoftwareX}\ }\textbf {\bibinfo {volume} {12}},\ \bibinfo {eid} {100620} (\bibinfo {year} {2020})},\ \Eprint {https://arxiv.org/abs/2007.11374} {arXiv:2007.11374 [astro-ph.IM]} \BibitemShut {NoStop}%
\bibitem [{\citenamefont {{Ashton}}\ \emph {et~al.}(2019)\citenamefont {{Ashton}} \emph {et~al.}}]{Ashton2019}%
  \BibitemOpen
  \bibfield  {author} {\bibinfo {author} {\bibfnamefont {G.}~\bibnamefont {{Ashton}}} \emph {et~al.},\ }\bibfield  {title} {\bibinfo {title} {{BILBY: A User-friendly Bayesian Inference Library for Gravitational-wave Astronomy}},\ }\href {https://doi.org/10.3847/1538-4365/ab06fc} {\bibfield  {journal} {\bibinfo  {journal} {Astrophys. J., Supp.}\ }\textbf {\bibinfo {volume} {241}},\ \bibinfo {eid} {27} (\bibinfo {year} {2019})},\ \Eprint {https://arxiv.org/abs/1811.02042} {arXiv:1811.02042 [astro-ph.IM]} \BibitemShut {NoStop}%
\bibitem [{\citenamefont {He}\ \emph {et~al.}(2015)\citenamefont {He}, \citenamefont {Zhang}, \citenamefont {Ren},\ and\ \citenamefont {Sun}}]{He2015}%
  \BibitemOpen
  \bibfield  {author} {\bibinfo {author} {\bibfnamefont {K.}~\bibnamefont {He}}, \bibinfo {author} {\bibfnamefont {X.}~\bibnamefont {Zhang}}, \bibinfo {author} {\bibfnamefont {S.}~\bibnamefont {Ren}},\ and\ \bibinfo {author} {\bibfnamefont {J.}~\bibnamefont {Sun}},\ }\href@noop {} {\bibinfo {title} {Deep residual learning for image recognition}} (\bibinfo {year} {2015}),\ \Eprint {https://arxiv.org/abs/1512.03385} {arXiv:1512.03385 [cs.CV]} \BibitemShut {NoStop}%
\bibitem [{\citenamefont {Abadi}\ \emph {et~al.}(2015)\citenamefont {Abadi} \emph {et~al.}}]{tensorflow2015-whitepaper}%
  \BibitemOpen
  \bibfield  {author} {\bibinfo {author} {\bibfnamefont {M.}~\bibnamefont {Abadi}} \emph {et~al.},\ }\href {https://www.tensorflow.org/} {\bibinfo {title} {{TensorFlow}: Large-scale machine learning on heterogeneous systems}} (\bibinfo {year} {2015}),\ \bibinfo {note} {software available from tensorflow.org}\BibitemShut {NoStop}%
\bibitem [{\citenamefont {{Sch{\"a}fer}}\ \emph {et~al.}(2022)\citenamefont {{Sch{\"a}fer}}, \citenamefont {{Zelenka}}, \citenamefont {{Nitz}}, \citenamefont {{Ohme}},\ and\ \citenamefont {{Br{\"u}gmann}}}]{Schafer2022a}%
  \BibitemOpen
  \bibfield  {author} {\bibinfo {author} {\bibfnamefont {M.~B.}\ \bibnamefont {{Sch{\"a}fer}}}, \bibinfo {author} {\bibfnamefont {O.}~\bibnamefont {{Zelenka}}}, \bibinfo {author} {\bibfnamefont {A.~H.}\ \bibnamefont {{Nitz}}}, \bibinfo {author} {\bibfnamefont {F.}~\bibnamefont {{Ohme}}},\ and\ \bibinfo {author} {\bibfnamefont {B.}~\bibnamefont {{Br{\"u}gmann}}},\ }\bibfield  {title} {\bibinfo {title} {{Training strategies for deep learning gravitational-wave searches}},\ }\href {https://doi.org/10.1103/PhysRevD.105.043002} {\bibfield  {journal} {\bibinfo  {journal} {\prd}\ }\textbf {\bibinfo {volume} {105}},\ \bibinfo {eid} {043002} (\bibinfo {year} {2022})},\ \Eprint {https://arxiv.org/abs/2106.03741} {arXiv:2106.03741 [astro-ph.IM]} \BibitemShut {NoStop}%
\bibitem [{\citenamefont {Kingma}\ and\ \citenamefont {Ba}(2017)}]{Kingma2017}%
  \BibitemOpen
  \bibfield  {author} {\bibinfo {author} {\bibfnamefont {D.~P.}\ \bibnamefont {Kingma}}\ and\ \bibinfo {author} {\bibfnamefont {J.}~\bibnamefont {Ba}},\ }\href@noop {} {\bibinfo {title} {Adam: A method for stochastic optimization}} (\bibinfo {year} {2017}),\ \Eprint {https://arxiv.org/abs/1412.6980} {arXiv:1412.6980 [cs.LG]} \BibitemShut {NoStop}%
\bibitem [{\citenamefont {{Koloniari}}\ \emph {et~al.}(2024)\citenamefont {{Koloniari}}, \citenamefont {{Koursoumpa}}, \citenamefont {{Nousi}}, \citenamefont {{Lampropoulos}}, \citenamefont {{Passalis}}, \citenamefont {{Tefas}},\ and\ \citenamefont {{Stergioulas}}}]{Koloniari2024}%
  \BibitemOpen
  \bibfield  {author} {\bibinfo {author} {\bibfnamefont {A.~E.}\ \bibnamefont {{Koloniari}}}, \bibinfo {author} {\bibfnamefont {E.~C.}\ \bibnamefont {{Koursoumpa}}}, \bibinfo {author} {\bibfnamefont {P.}~\bibnamefont {{Nousi}}}, \bibinfo {author} {\bibfnamefont {P.}~\bibnamefont {{Lampropoulos}}}, \bibinfo {author} {\bibfnamefont {N.}~\bibnamefont {{Passalis}}}, \bibinfo {author} {\bibfnamefont {A.}~\bibnamefont {{Tefas}}},\ and\ \bibinfo {author} {\bibfnamefont {N.}~\bibnamefont {{Stergioulas}}},\ }\bibfield  {title} {\bibinfo {title} {{New Gravitational Wave Discoveries Enabled by Machine Learning}},\ }\href {https://doi.org/10.48550/arXiv.2407.07820} {\bibfield  {journal} {\bibinfo  {journal} {arXiv e-prints}\ ,\ \bibinfo {eid} {arXiv:2407.07820}} (\bibinfo {year} {2024})},\ \Eprint {https://arxiv.org/abs/2407.07820} {arXiv:2407.07820 [gr-qc]} \BibitemShut {NoStop}%
\bibitem [{\citenamefont {{Sch{\"a}fer}}\ and\ \citenamefont {{Nitz}}(2022)}]{Schafer2022b}%
  \BibitemOpen
  \bibfield  {author} {\bibinfo {author} {\bibfnamefont {M.~B.}\ \bibnamefont {{Sch{\"a}fer}}}\ and\ \bibinfo {author} {\bibfnamefont {A.~H.}\ \bibnamefont {{Nitz}}},\ }\bibfield  {title} {\bibinfo {title} {{From one to many: A deep learning coincident gravitational-wave search}},\ }\href {https://doi.org/10.1103/PhysRevD.105.043003} {\bibfield  {journal} {\bibinfo  {journal} {\prd}\ }\textbf {\bibinfo {volume} {105}},\ \bibinfo {eid} {043003} (\bibinfo {year} {2022})},\ \Eprint {https://arxiv.org/abs/2108.10715} {arXiv:2108.10715 [astro-ph.IM]} \BibitemShut {NoStop}%
\bibitem [{\citenamefont {{Abbott}}\ \emph {et~al.}(2023)\citenamefont {{Abbott}} \emph {et~al.}}]{GWTC3-data}%
  \BibitemOpen
  \bibfield  {author} {\bibinfo {author} {\bibfnamefont {R.}~\bibnamefont {{Abbott}}} \emph {et~al.} (\bibinfo {collaboration} {{LIGO Scientific Collaboration} and {Virgo Collaboration} and {KAGRA Collaboration}}),\ }\bibfield  {title} {\bibinfo {title} {{GWTC-3: Compact Binary Coalescences Observed by LIGO and Virgo During the Second Part of the Third Observing Run — O3 search sensitivity estimates}},\ }\href {https://doi.org/10.5281/zenodo.7890437} {10.5281/zenodo.7890437} (\bibinfo {year} {2023})\BibitemShut {NoStop}%
\bibitem [{\citenamefont {{Ewing}}\ \emph {et~al.}(2024)\citenamefont {{Ewing}} \emph {et~al.}}]{Ewing2024}%
  \BibitemOpen
  \bibfield  {author} {\bibinfo {author} {\bibfnamefont {B.}~\bibnamefont {{Ewing}}} \emph {et~al.},\ }\bibfield  {title} {\bibinfo {title} {{Performance of the low-latency GstLAL inspiral search towards LIGO, Virgo, and KAGRA's fourth observing run}},\ }\href {https://doi.org/10.1103/PhysRevD.109.042008} {\bibfield  {journal} {\bibinfo  {journal} {\prd}\ }\textbf {\bibinfo {volume} {109}},\ \bibinfo {eid} {042008} (\bibinfo {year} {2024})},\ \Eprint {https://arxiv.org/abs/2305.05625} {arXiv:2305.05625 [gr-qc]} \BibitemShut {NoStop}%
\bibitem [{\citenamefont {{Usman}}\ \emph {et~al.}(2016)\citenamefont {{Usman}} \emph {et~al.}}]{Usman2016}%
  \BibitemOpen
  \bibfield  {author} {\bibinfo {author} {\bibfnamefont {S.~A.}\ \bibnamefont {{Usman}}} \emph {et~al.},\ }\bibfield  {title} {\bibinfo {title} {{The PyCBC search for gravitational waves from compact binary coalescence}},\ }\href {https://doi.org/10.1088/0264-9381/33/21/215004} {\bibfield  {journal} {\bibinfo  {journal} {Classical and Quantum Gravity}\ }\textbf {\bibinfo {volume} {33}},\ \bibinfo {eid} {215004} (\bibinfo {year} {2016})},\ \Eprint {https://arxiv.org/abs/1508.02357} {arXiv:1508.02357 [gr-qc]} \BibitemShut {NoStop}%
\bibitem [{\citenamefont {Farr}(2019)}]{Farr2019}%
  \BibitemOpen
  \bibfield  {author} {\bibinfo {author} {\bibfnamefont {W.~M.}\ \bibnamefont {Farr}},\ }\bibfield  {title} {\bibinfo {title} {Accuracy requirements for empirically measured selection functions},\ }\href {https://doi.org/10.3847/2515-5172/ab1d5f} {\bibfield  {journal} {\bibinfo  {journal} {Research Notes of the AAS}\ }\textbf {\bibinfo {volume} {3}},\ \bibinfo {pages} {66} (\bibinfo {year} {2019})}\BibitemShut {NoStop}%
\bibitem [{GWS()}]{GWSamplegen}%
  \BibitemOpen
  \href@noop {} {}\bibinfo {howpublished} {\url{https://github.com/alistair-mcleod/GWSamplegen}}\BibitemShut {NoStop}%
\bibitem [{Inf()}]{Infernus}%
  \BibitemOpen
  \href@noop {} {}\bibinfo {howpublished} {\url{https://github.com/alistair-mcleod/infernus}}\BibitemShut {NoStop}%
\bibitem [{GWO()}]{GWOSC_site}%
  \BibitemOpen
  \href@noop {} {}\bibinfo {howpublished} {\url{https://gwosc.org}}\BibitemShut {NoStop}%
\end{thebibliography}%

\end{document}